\documentclass[a4paper,10pt]{article}
\usepackage[margin=3cm]{geometry}
\usepackage{graphicx}
\usepackage{dsfont,amsmath,amssymb}

\newcommand{\vev}[1]{\langle #1 \rangle}
\newcommand{\e}{\mathrm{e}}

\newcommand \be  {\begin{equation}}
\newcommand \bea {\begin{eqnarray} \nonumber }
\newcommand \ee  {\end{equation}}
\newcommand \eea {\end{eqnarray}}

\newcommand{\tr}{\operatorname{Tr}}

\renewcommand{\d     }[1]{\mathrm{d}#1} % pour les intégrales
  \newcommand{\1     }[1]{\mathds{1}_{\{#1\}}}

\title{The fine-structure of volatility feedback I:\\ multi-scale self-reflexivity}
\author{R\'emy~Chicheportiche$^{1,2}$ and Jean-Philippe~Bouchaud$^1$\\[15pt]
{\normalsize  $^1$ Capital~Fund~Management, 75\,009 Paris, France}\\
{\normalsize  $^2$ Chaire de finance quantitative, Ecole Centrale Paris, 92\,295 Ch\^atenay-Malabry, France}
}
\date{\today}
\begin{document}

\maketitle

\begin{abstract}

We attempt to unveil the fine structure of volatility feedback effects in the context of general quadratic autoregressive (QARCH) models,
which assume that today's volatility can be expressed as a general quadratic form of the past daily returns. 
The standard ARCH or GARCH framework is recovered when the quadratic kernel is diagonal. 
The calibration of these models on US stock returns reveals several unexpected features. 
The off-diagonal (non ARCH) coefficients of the quadratic kernel are found to be highly significant both In-Sample and Out-of-Sample, 
although all these coefficients turn out to be one order of magnitude smaller than the diagonal elements. 
This confirms that daily returns play a special role in the volatility feedback mechanism, as postulated by ARCH models. 
The feedback kernel exhibits a surprisingly complex structure, incompatible with all models proposed so far in the literature.
Its spectral properties suggest the existence of volatility-neutral patterns of past returns. 
The diagonal part of the quadratic kernel is found to decay as a power-law of the lag, in line with the long-memory of volatility. 
Finally, QARCH models suggest some violations of Time Reversal Symmetry in financial time series, 
which are indeed observed empirically, although of much smaller amplitude than predicted. 
We speculate that a faithful volatility model should include both ARCH feedback effects and a stochastic component. 
\end{abstract}

%%%%%%%%%%%%%%%%%%%%%%%%%%%%%%%%%%%%%%%%%%%%%%%%%%%%%%%%%%%%%%%%%%%%%%%%

%%%%%%%%%%%%%%%%%%%%%%%%%%%%%%%%%%%%%%%%%%%%%%%%%%%%%%%%%%%%%%%%%%%%%%%%
\section{Introduction}

One of the most striking universal stylized facts of financial returns is the volatility clustering effect,
which was first reported by Mandelbrot as early as 1963 \cite{mandelbrot1963variation}. He noted that \ldots
{\it large changes tend to be followed by large changes, of either sign, and small changes tend to be followed by small changes.}
The first quantitative description of this effect was the ARCH model proposed by Engle in 1982 \cite{engle1982autoregressive}. 
It formalizes Mandelbrot's hunch in the simplest possible way, 
by postulating that returns $r_t$ are conditionally Gaussian random variables, 
with a time dependent volatility (rms) $\sigma_t$ that evolves according to:
\be
\sigma^2_t = s^2 + g r_{t-1}^2.
\ee
In words, this equation means that the (squared) volatility today is equal to a baseline level $s^2$, 
plus a self-exciting term that describes the influence of yesterday's {\it perceived} volatility $r_{t-1}^2$ on today's activity, 
through a feedback parameter $g$. 
Note that this ARCH model was primarily thought of as an econometric model that needs to be calibrated on data, 
while a more ambitious goal would be to {\it derive} such a model from a more fundamental theory 
--- for example, based on behavioural reactions to perceived risk. 

It soon became apparent that the above model is far too simple to account for empirical data. 
For one thing, it is unable to account for the long memory nature of volatility fluctuations. 
It is also arbitrary in at least two ways:
\begin{itemize} 

\item First, there is no reason to limit the feedback effect to the previous day only. 
      The Generalized ARCH model (GARCH) \cite{bollerslev1986generalized}, which has become a classic in quantitative finance, 
      replaces $r_{t-1}^2$ by an exponential moving average of past squared returns. 
      Obviously, one can further replace the exponential moving average by any weighting kernel $k(\tau)\geq 0$, 
      leading to a large family of models such as:
      \be
	\sigma^2_t = s^2 + \sum_{\tau=1}^\infty k(\tau) r_{t-\tau}^2,
      \ee
      which includes all ARCH and GARCH models. 
      For example, ARCH($q$) corresponds to a kernel $k(\tau)$ that is strictly zero beyond $\tau=q$. 
      A slowly (power-law) decaying kernel $k(\tau)$ is indeed able to account for the long memory of volatility 
      --- this corresponds to the so-called FIGARCH model (for Fractionally Integrated GARCH) \cite{bollerslev1994arch}. 

\item Second, there is no {\it a priori} reason to single out the day as the only time scale to define the returns:
      in principle, returns over different time scales  could also feedback on the volatility today 
      \cite{muller1997volatilities,borland2005multi,lynch2003market}.
      For extended time scales \emph{longer} than the day, this
      leads to another natural extension of the GARCH model as:
      \be
	\sigma^2_t = s^2 + \sum_{\ell} \sum_{\tau=1}^\infty g_\ell(\tau) R^{(\ell)2}_{t-\tau},
      \ee
      where $R^{(\ell)}_t$ is the cumulative, $\ell$ day return between $t-\ell$ and $t$. 
      The first model in that category is the HARCH model of the Olsen group \cite{muller1997volatilities}, 
      where the first ``H'' stands for Heterogeneous. 
      The authors had in mind that different traders are sensitive to and react to returns on different time scales.
      Although this behavioural interpretation was clearly expressed, 
      there has been no real attempt%
      \footnote{See however the very recent stochastic volatility model with heterogeneous time scales of \cite{delpini2012stochastic}.} 
      to formalize such an intuition beyond the hand-waving arguments given in \cite{borland2005multi}.
      {As of the subdaily scales, we study in a companion paper \cite{Pierre_inprep} the respective feedbacks of overnight and day returns onto both overnight and day volatilities.}
\end{itemize}

The common point to the zoo of generalizations of the initial ARCH model is that 
the current level of volatility $\sigma_t^2$ is expressed as a quadratic form of past realized returns. 
The most general model of this kind, called QARCH (for Quadratic ARCH), is due to Sentana \cite{sentana1995quadratic}, and reads:
\be\label{eq:quadraticARCH}
\sigma^2_t = s^2 + \sum_{\tau=1}^\infty L(\tau) \, r_{t-\tau} + \sum_{\tau,\tau'=1}^\infty K(\tau,\tau') \, r_{t-\tau}r_{t-\tau'},
\ee
where $L(\tau)$ and $K(\tau,\tau')=K(\tau',\tau)$ are some kernels that should satisfy technical conditions for $\sigma^2_t$ to be always positive 
(see below and \cite{sentana1995quadratic}).
The QARCH can be seen as a general discrete-time model for the dependence of $\sigma_t^2$ on all past returns $\left\{r_{t'}\right\}_{t'<t}$, truncated to second order. 
The linear contribution, which involves $L(\tau)$, captures a possible dependence of the volatility on the sign of the past returns. 
For example, negative past returns tend to induce larger volatility in the future 
--- this is the well-known leverage effect \cite{black1976studies,bouchaud2001leverage,bekaert2000asymmetric}, see also \cite{reigneron2011principal} and references therein.%
\footnote{(G)QARCH and alternative names such as Asymmetric (G)ARCH, Nonlinear (G)ARCH, Augmented ARCH, etc.\ 
often refer to this additional leverage (asymmetry) contribution,
whereas the important innovation of QARCH is in fact the possibility of off-diagonal terms in the kernel $K$.} 
The quadratic contribution, on the other hand, contains through the matrix $K(\tau,\tau')$ all ARCH models studied in the literature. 
For example, ARCH($q$), GARCH and FIGARCH models all correspond to a purely {\it diagonal} kernel, $K(\tau,\tau')=k(\tau) \delta_{\tau,\tau'}$, 
where $\delta_{\tau,\tau'}$ is Kronecker's delta. 

In view of the importance of ARCH modelling in finance, it is somewhat surprising that the general framework provided by QARCH has not been fully explored. 
Only versions with very short memories, corresponding to at most $2 \times 2$ matrices for $K$, seem to have been considered in the literature. 
In fact, Sentana's contribution is usually considered to be the introduction of the linear contribution in the GARCH framework, 
rather than unveiling the versatility of the quadratic structure of the model. 
The aim of the present paper is to explore in detail the QARCH framework, both from a theoretical and empirical point of view. 
Of particular interest is the empirical determination of the structure of the feedback kernel $K(\tau,\tau')$ for the daily returns of stocks, 
which we compare with several proposals in the literature, 
including the multiscale model of \cite{borland2005multi} and the trend-induced volatility model of \cite{zumbach2010volatility}. 
Quite surprisingly, we find that while the off-diagonal elements of $K(\tau,\tau')$ are significant, 
they are at least an order of magnitude smaller than the diagonal elements $k(\tau) := K(\tau,\tau)$.  
The latter are found to decay very slowly with $\tau$, in agreement with previous discussions. 
Therefore, in a first approximation, the dominant feedback effect comes from the amplitude of {\it daily returns} only, 
with minor corrections coming from returns computed on large time spans, at variance with the assumption of the model 
put forward in \cite{borland2005multi}. We believe that this finding is unexpected and far from trivial. 
It is a strong constraint on any attempt to justify the ARCH feedback mechanism from a more fundamental point of view.
{On the other hand, important corrections to the pure (daily close-to-close) GARCH structure are also found to stem from the interplay of 
past open-to-close and close-to-open returns, as well as their distinct feedback in the future volatility \cite{Pierre_inprep}.}

In parallel with ARCH modelling,  stochastic volatility models represent another strand of the literature that has vigorously grown in the last twenty years. 
Here again, a whole slew of models has emerged \cite{henry2008analysis}, with the Heston model \cite{heston1993closed} and the SABR model \cite{hagan2002managing} as the best known examples. 
These models assume that the volatility itself is a random process, governed either by a stochastic differential equation (in time) 
or an explicit cascade construction in the case of more recent multifractal models \cite{bacrymuzy,calvetfisher,lux}
(again initiated by Mandelbrot as early as 1974! \cite{mandelbrot1974intermittent}). 
There is however a fundamental difference between most of these stochastic volatility models and the ARCH framework: 
while the former are {\it time-reversal invariant} (TRI), the latter is explicitly {\it backward looking}. 
This, as we shall discuss below, implies that certain correlation functions are not TRI within QARCH models, 
but are TRI within stochastic volatility models. 
This leads to an empirically testable prediction; we report below that TRI is indeed violated in stock markets, as also documented in \cite{zumbach2009time}. 

The outline of this paper is as follows. 
We first review in Section~2 some general analytical properties of QARCH models, 
in particular about the existence of low moments of the volatility. 
We then introduce in Section~3 several different sub-families of QARCH, that we try to motivate intuitively. 
The consideration of these sub-families follows from the necessity of reducing the dimensionality of the problem, 
but also from the hope of finding simple regularities that would suggest a plausible interpretation (behavioural or else) of the model, 
beyond merely best fit criteria. 
In Section~4, we attempt to calibrate ``large'' QARCH models on individual stock returns, 
first without trying to impose any a priori structure on the kernel $K(\tau,\tau')$, 
and then specializing to the various sub-families mentioned above. 
We isolate in Section~5 the discussion on the issue of TRI for stock returns, both from a theoretical/modeling and an empirical point of view.
We give our conclusions in Section~6, and relegate to appendices more technical issues, and the calibration of the model on the returns of the 
stock index.

\section{General properties of QARCH models}\label{sec:section2}

Some general properties of QARCH models are discussed in Sentana's seminal paper \cite{sentana1995quadratic}. 
We review them here and derive some new results. 
The QARCH model for the return at time $t$, $r_t$, is such that:
\be
\ln p_{t} - \ln p_{t-1} = r_t = \sigma_t \xi_t,
\ee
where $p_t$ is the price at time $t$, $\sigma_t$ is given by the QARCH specification, Eq.~\eqref{eq:quadraticARCH} above, 
while the $\xi$'s are IID random variables, of zero mean and variance equal to unity. 
While many papers take these $\xi$'s to be Gaussian, it is preferable to be agnostic about their univariate distribution. 
In fact, several studies including our own (see below), suggest that the $\xi$'s themselves have fat-tails: 
asset returns are {\it not} conditionally Gaussian and ``true jumps'' do occur.\footnote{There seems to be a slowly growing
consensus on this point (see e.g.\ \cite{ait2009analyzing}): Gaussian processes with stochastic volatility cannot alone account for 
the discontinuities observed in market prices.}

In this section, we will focus on the following non-linear correlation functions 
(other correlations will be considered in Appendix~\ref{sec:first_est}, when we turn to empirical studies):
\begin{subequations}
\begin{align}
	           \mathcal{C}^{(2)}(\tau)   &\equiv\vev{\left(r^2_t-\vev{r_{t'}^2}_{t'}\right)             r^2_{t-\tau}}_t\\
	\widetilde{\mathcal{C}}^{(2)}(\tau)  &\equiv\vev{\left(\sigma^2_t-\vev{\sigma_{t'}^2}_{t'}\right)   r^2_{t-\tau}}_t\\
	           \mathcal{D} (\tau',\tau'')&\equiv\vev{\left((r^2_{t}-\vev{r_{t'}^2}_{t'}\right)          r_{t-\tau'}r_{t-\tau''}}_t\\
	\widetilde{\mathcal{D}}(\tau',\tau'')&\equiv\vev{\left((\sigma^2_{t}-\vev{\sigma_{t'}^2}_{t'}\right)r_{t-\tau'}r_{t-\tau''}}_t.
\end{align}
\end{subequations}
Here and below, we assume stationarity and correspondingly $\vev{\dots}_t$ refers to a sliding average over $t$. 
The following properties are worth noticing: by definition, 
$\mathcal{D}(\tau,\tau) \equiv \mathcal{C}^{(2)}(\tau)$ and 
$\widetilde{\mathcal{D}}(\tau,\tau) \equiv \widetilde{\mathcal{C}}^{(2)}(\tau)$.
Furthermore, whereas $\mathcal{C}^{(2)}(\tau) = \mathcal{C}^{(2)}(-\tau)$ by construction, 
the same is not true in general for $\widetilde{\mathcal{C}}^{(2)}(\tau)$. 
However, using the QARCH causal construction and the independence of the $\xi$'s, 
one can easily convince oneself that when $\tau > 0$, $\widetilde{\mathcal{C}}^{(2)}(\tau) \equiv \mathcal{C}^{(2)}(\tau)$. 
Similarly, for \mbox{$\tau'>\tau''>0$}, $\widetilde{\mathcal{D}}(\tau',\tau'') \equiv \mathcal{D}(\tau',\tau'')$, 
while in general, $\mathcal{D}(\tau',\tau'') \neq \mathcal{D}(-\tau',-\tau'') \equiv 0$. 

%\subsection{Second moment of the volatility and stationarity}

\begin{figure}
	\center
	\includegraphics[scale=0.55,trim=  0 0 890 0,clip]{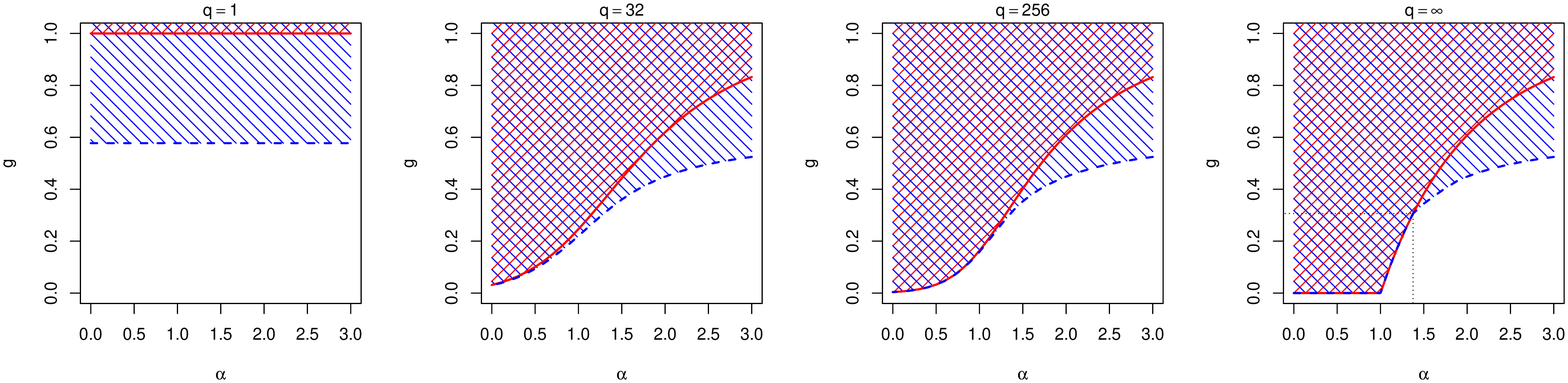}
	\includegraphics[scale=0.55,trim=315 0 600 0,clip]{critfront24_allq.eps}
	\includegraphics[scale=0.55,trim=890 0   0 0,clip]{critfront24_allq.eps}
	\caption{Allowed region in the $\alpha,g$ space for $K(\tau,\tau)=g\,\tau^{-\alpha}\1{\tau\leq q}$ and $L(\tau)=0$,
	         according to the finiteness of $\vev{\sigma^2}$ and $\vev{\sigma^4}$. 
             Divergence of $\vev{\sigma^2}$ is depicted by  $45^\circ$ (red)  hatching, while 
             divergence of $\vev{\sigma^4}$ is depicted by $-45^\circ$ (blue) hatching.
             In the wedge between the dashed blue and solid red lines, 
             \mbox{$\vev{\sigma^2} < \infty$} while $\vev{\sigma^4}$ diverges.}
	\label{fig:sig2crit}
\end{figure}

QARCH models only make sense if the expected volatility does not diverge to infinity. 
The criterion for stability is easy to establish if the $\xi$'s are IID and of zero mean, and reads:
\be
\tr K < 1.
\ee
In this case, the volatility is a stationary process such that $\vev{\sigma^2}\equiv\mathds{E}[\sigma^2]=s^2/(1 - \tr K)$: 
the feedback-induced increase of the volatility only involves the diagonal elements of $K$. 
Note also that the leverage kernel $L(\tau)$ does not appear in this equation. 
As an interesting example, we consider kernels with a power-law decaying diagonal: 
$K(\tau,\tau)=g\,\tau^{-\alpha}\mathds{1}_{\{\tau\leq q\}}$. 
For a given $\alpha$, $g$ must be smaller than a certain $g_\text{c}(\alpha,q)$ for $\vev{\sigma^2}$ to be finite. 
Fig.~\ref{fig:sig2crit} shows the critical frontier $g_\text{c}(\alpha,q)$ for $q=1,32,256$ and $q \to \infty$. 
The critical frontier in the limit case $q=\infty$ is given by $g_\text{c}=1/\zeta(\alpha)$, 
where $\zeta(\alpha)$ is Riemann's zeta function (solid red). 
Note in particular that the model is always unstable when \mbox{$\alpha < 1$}, 
i.e.\ when the memory of past realized volatility decays too slowly.%
\footnote{In the context of fractionally integrated processes $I(d)$, the condition $\alpha\leq 1$
is equivalent to the `difference parameter' $d=\alpha-1$ being positive.} 
At the other extreme, $q=1$, the constraint is well known to be $g=k(1)\leq 1$ (solid red).

Within a strict interpretation of the QARCH model, there are additional constraints on the kernels $K$ and $L$ 
that arise from the fact that $\sigma_t^2$ should be positive for any realization of price returns. 
This imposes that a) all the eigenvalues of $K$ should be non-negative, and b) that the following inequality holds:
\be\label{eq:nonneg}
\sum_{\tau,\tau'=1}^q L(\tau) K^{-1}(\tau,\tau') L(\tau') \leq 4\,s^2,
\ee
where $K^{-1}$ is the matrix inverse of $K$. 
However, these constraints might be too strong if one interprets the QARCH model as a generic expansion of $\sigma^2_t$ in powers of past returns,
truncated to second order \cite{sentana1995quadratic}. 
It could well be that higher order terms are stabilizing and lead to a meaningful, stable model beyond the limits quoted here. Still, the empirically
calibrated model will be found to satisfy the above inequality by quite a large margin.

The existence of higher moments of $\sigma$ can also be analyzed, leading to more and more cumbersome algebra 
\cite{he1999fourth,ling2002stationarity,teyssiere2010long}. 
In view of its importance, we have studied in detail in Appendix~\ref{sec:fourth_moment} the conditions for the existence of the fourth moment of $\sigma$, 
which allows one to characterize the excess kurtosis $\kappa$ of the returns, traditionally defined as:
\be
\kappa = \frac{\langle r^4 \rangle}{\langle r^2 \rangle^2} - 3 \equiv \frac{\langle \sigma^4 \rangle \langle \xi^4 \rangle}{\langle \sigma^2 \rangle^2} - 3.
\ee

We show in particular that for a FIGARCH with $K(\tau,\tau)=g\tau^{-\alpha}$ and $L(\tau)=0$, 
$\vev{\sigma^4}$ only diverges if \mbox{$\alpha < 1/2$}, 
but this is far in the forbidden region \mbox{$\alpha < 1$} where $\langle \sigma^2 \rangle$ itself diverges, see Fig.~\ref{fig:sig2crit}. 
Therefore, a FIGARCH model with a long memory (i.e.\ \mbox{$\alpha < 1.376$}) cannot lead to a 
large kurtosis of the returns, {\it unless the $\xi$ variables have themselves fat tails}. 
We will come back to this important point below.

An important remark must be made at this stage: as we demonstrate in Appendix~\ref{QARCHapx:B}, 
the ubiquitous empirical long-memory correlations of the volatility $\mathcal{C}^{(2)}(\tau)\propto \tau^{-\beta}$ with 
$0 < \beta < 1$ can be reproduced by FIGARCH models with $\alpha= (3 - \beta)/2 \in (1,3/2)$.

\section{Some special families of QARCH models}\label{sec:section3}
\begin{figure}
	\center
	\includegraphics[scale=0.5]{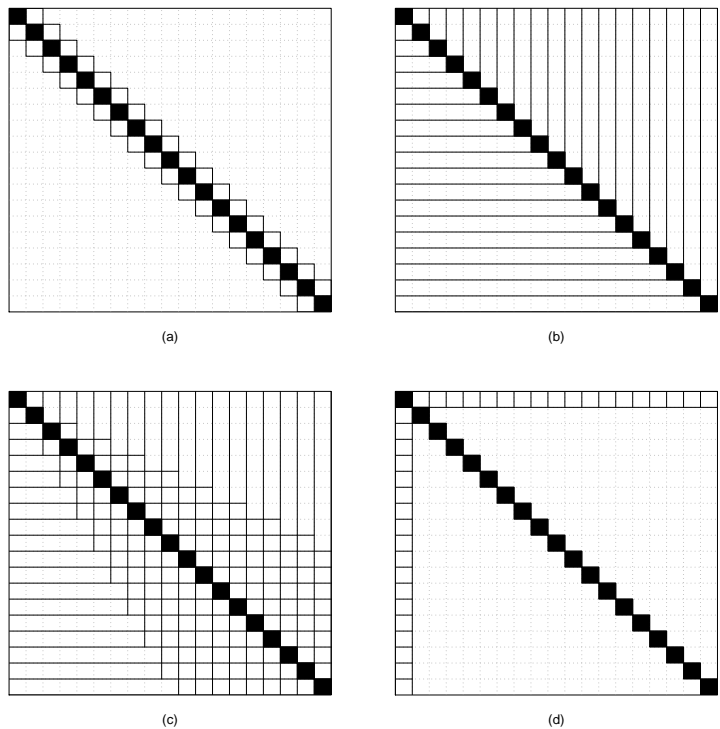}
	\caption{Some simple kernel structures. \textbf{(a)} Overlapping two-scales; \textbf{(b)} Borland-Bouchaud multiscale; 
	                            \textbf{(c)} Zumbach; \textbf{(d)} Long-Trend.}
	\label{fig:matrices}
\end{figure}

\paragraph{}
As we alluded to in the introduction, ARCH($q$) models posit that today's volatility is only sensitive to past daily returns. This assumption can be relaxed in several natural ways,
each of which leading to a specific structure of the feedback kernel $K$. We will present these extensions in increasing order of complexity.  

\subsection{Returns over different time scales}

Let us define the $\ell$-day return between $t-\ell$ and $t$ as $R^{(\ell)}_t$, such that:
\be
R^{(\ell)}_t = \sum_{\tau=1}^{\ell} r_{t-\tau}; \qquad R^{(1)}_t = r_{t-1} = \ln p_{t-1} - \ln p_{t-2},
\ee
where $p_t$ is the price at time $t$. The simplest extension of ARCH($q$) is to allow all past 2-day returns to play a role as well, i.e.:
\be
\sigma_t^2 = s^2 + \sum_{\tau=0}^{q-1} \, g_1(\tau) [R^{(1)}_{t-\tau}]^2 + \sum_{\tau=0}^{q-2} \, g_2(\tau) [R^{(2)}_{t-\tau}]^2,
\ee
where $g_1(\tau)$ and $g_2(\tau)$ are coefficients, all together $2q-1$ of them. Upon identification with the QARCH kernel, one finds:
\begin{align}\nonumber
K(\tau,\tau)   &=  g_1(\tau-1) + g_2(\tau-1) + g_2(\tau-2),\\
K(\tau,\tau+1) &= g_2(\tau-1)\\\nonumber
K(\tau,\tau+\ell) &= 0 \, {\text{ for }}\, \ell \geq 2,
\end{align}
with the convention that $g_2(-1)=0$. The model can thus be re-interpreted in the following way: 
the square volatility is still a weighted sum of past {\it daily} squared returns, 
but there is an extra contribution that picks up the realized one-day covariance of returns. 
If $g_2(\tau) \geq 0$, the model means that the persistence of the same trend two days in a row leads to increased volatilities. 
A schematic representation of this model is given in Fig.~\ref{fig:matrices}(a).

One can naturally generalize again the above model to include 2-day, 3-day, $\ell$-day returns, 
with more coefficients $g_1(\tau), g_2(\tau), \dots, g_{\ell}(\tau)$, with a total of $\ell(2q+1-\ell)/2$ parameters. 
Obviously, when $\ell=q$, all possible time scales are exhausted, and the number of free parameters is $q(q+1)/2$, 
i.e.\ exactly the number of independent entries of the symmetric $q \times q$ feedback kernel $K$. 

\subsection{Multiscale, cumulative returns}

Another model, proposed in \cite{muller1997volatilities,borland2005multi}, is motivated by the idea that traders may be influenced not only by yesterday's return, but also by the change of 
price between today and 5-days ago (for weekly traders), or today and 20-days ago (for monthly traders), etc. In this case, the natural extension of the ARCH framework is to 
write:
\be
\sigma^2_t = s^2 + \sum_{\ell=1}^q g_{\text{BB}}(\ell) [R^{(\ell)}_{t}]^2,
\ee
where the index BB refers to the model put forward in \cite{borland2005multi}. The BB model requires a priori $q$ different parameters. 
It is simple to see that in this case, the kernel matrix can be expressed as:
\be
K_{\text{BB}}(\tau',\tau'') = G[\max(\tau',\tau'')], \quad {\mbox{with}} \quad G[\tau] = \sum_{\ell=\tau}^q g_{\text{BB}}(\ell).
\ee
The spectral properties of these matrices are investigated in detail in \cite{chicheportiche2013phd}. % Appendix~\ref{apx:A}.
One can also consider a mixed model where both cumulative returns and daily returns play a role. 
This amounts to taking the off-diagonal elements of $K$ as prescribed by the above equation, 
but to specify the diagonal elements $K(\tau,\tau)$ completely independently from $G[\tau]$. 
This leads to a matrix structure schematically represented in Fig.~\ref{fig:matrices}(b), parameterized by $2q-1$ independent coefficients. 

\subsection{Zumbach's trend effect (ARTCH)}

Zumbach's model \cite{zumbach2010volatility} is another particular case in the class of models described by Eq.~\eqref{eq:quadraticARCH}.
It involves returns over different lengths of time and characterizes the effect of past trending aggregated returns on future volatility.
The quadratic part in the volatility prediction model is
\be
	\text{ARCH}+\sum_{\ell= 1}^{\lfloor{q/2\rfloor}} g_{\text{Z}}({\ell})\,R^{(\ell)}_{t}R^{(\ell)}_{t\!-\!\ell}.
\ee
When relevant, only specific time scales like the day ($\ell=1$), the week ($\ell= 5$), the month ($\ell = 20$), etc.\ can be retained in the summation.
The off-diagonal elements of the kernel $K$ now take the following form:
\begin{equation}\label{eq:K.Z}
	K_{\text{Z}}(\tau',\tau'' > \tau')=\sum_{\ell = \max(\tau',\tfrac{\tau''}{2})}^{\min(\tau''\!-\!1,\lfloor{q/2\rfloor})}g_{\text{Z}}({\ell})
\end{equation}
Since it is upper triangular by construction, we first symmetrize it, and the diagonal is filled with the ARCH parameters.
This model contains $q+ \lfloor{q/2\rfloor}$ independent coefficients, and is schematically represented in Fig.~\ref{fig:matrices}(c). 

\subsection{A generalized trend effect}

In Zumbach's model, the trend component is defined by comparing returns computed over the same horizon $\ell$. 
This of course is not necessary. 
As an extreme alternative model, we consider a model where the volatility today is affected by 
the last return $r_{t-1}$ confirmation (or the negation) of a long trend.
In more formal terms, this writes:
\be
	\text{ARCH}+  r_{t-1} \times \sum_{\ell=1}^{q-1} g_{\text{LT}}({\ell})  r_{t-1-\ell}, 
\ee
where $g_{\text{LT}}(\ell)$ is the sequence of weights that defines the past ``long trend'' (hence the index LT). 
This now corresponds to a kernel $K$ with diagonal elements corresponding to the ARCH effects 
and a single non trivial row (and column) corresponding to the trend effect: $K(1,\tau>1) = g_{\text{LT}}(\tau-1)$.
This model has again $2q-1$ free parameters.

Of course, one can consider QARCH models that encode some, or all of the above mechanisms --- 
for example, a model that schematically reads $\text{ARCH} + \text{BB} + \text{LT}$ would require $3q-2$ parameters.

\subsection{Spectral interpretation of the QARCH}\label{sec:spectral}

Another illuminating way to interpret QARCH models is to work  
in the diagonal basis of the $K$ matrix, where the quadratic term in \eqref{eq:quadraticARCH} reads:
\begin{equation}\label{eq:eigen_decomp}
	 \sum_{\tau',\tau''=1}^{q}\left(\sum_{n}\lambda_nv_n(\tau')v_n(\tau'')\right)\,r_{t-\tau'}\,r_{t-\tau''}
	\equiv \sum_{n}\lambda_n\, \langle r | v_n \rangle_t^2
\end{equation}
with $(\lambda_n,v_n)$ being, respectively, the $n$-th eigenvalue and eigenvector of $K$, 
and $\langle r | v_n \rangle_t = \sum_{\tau=1}^{q}v_n(\tau)\,r_{t-\tau}$
 the projection of the pattern created by the last $q$ returns on the $n$-th eigenvector. 
One can therefore say that the square volatility $\sigma_t^2$ picks up contributions from various past returns eigenmodes. 
The modes associated to the largest eigenvalues $\lambda$ are those which have the largest contribution to volatility spikes. 

The ARCH($q$) model corresponds to a diagonal matrix; in this case the modes are trivially individual daily returns. 
Another trivial case is when $K$ is of rank one and its spectral decomposition is simply
\begin{equation}\label{eq:K.rank1}
	K(\tau',\tau'')=\lambda v(\tau')v(\tau'')
\end{equation}
where $\lambda=\tr(K)$ is the only non-null eigenvalue, 
and $v(\tau)=\sqrt{K(\tau,\tau)/\tr(K)}$ the eigenvector associated with this non-degenerate mode.
The corresponding contribution to the increase in volatility \eqref{eq:eigen_decomp} is therefore $\lambda \widehat R^2_t$,
where
\be
\widehat R_t = \vev{r|v}_t=\sum_{\tau=1}^qv(\tau)\,r_{t-\tau},
\ee
can be interpreted as an average return over the whole period, with a certain set of weights $v(\tau)$. 

The pure BB model (without extra ARCH contributions) can also be diagonalized analytically in the large $q$ limit for certain choices of the function $g_\text{BB}(\tau)$ \cite{chicheportiche2013phd}. 
%We detail these calculations (which are mostly of theoretical interest) in Appendix~\ref{apx:A}. 

\section{Empirical study: single stocks}

We now turn to the calibration on single stock returns  of general, ``large'' QARCH models, i.e.\ models that take into account $q$ past returns with $q$ large (20 or more). 
The difficulty is that the full calibration of the matrix $K$ requires the determination of $q(q+1)/2$ parameters, which is already 
equal to $210$ when $q=20$! Imposing some {\it a priori} structure (like the ones discussed in the previous section) on the
matrix $K$ may help limiting the number of parameters, and gaining robustness and transparency. However, perhaps surprisingly, we will find that
none of the above model seem to have enough flexibility to reproduce the subtle structure of the empirically determined $K$ matrix.

The standard procedure used to calibrate ARCH models is maximum-likelihood, which 
relies on the choice of a family of distributions for the noise component $\xi$, often taken to be Student-t 
distributions. However, this method cannot be used 
directly in our case because there are far too many parameters and the numerical optimization of the log-likelihood 
is both extremely demanding in computer time and unreliable, as many different optima are expected in general. 
An alternative method, the Generalized Method of Moments (GMM), is to determine the \mbox{$1+q+q(q+1)/2$} 
parameters of the model using empirically determined correlation functions that depend on $s^2$, $L(\tau)$ and $K(\tau,\tau')$ 
--- see equations~\eqref{eq:model_predictions} in Appendix~\ref{apx:methodology}.
This latter method is however sensitive to tail events and can lead to biases. 
We will therefore use a hybrid strategy, where a first estimate of these parameters, 
obtained using the GMM, serves as the starting point of a one-step likelihood maximization, 
which determines the set of most likely parameters in the vicinity of the GMM estimate 
(more details on this below).

The description of our data set, and various important methodological issues are discussed in Appendix~\ref{apx:methodology}.
An important point is that we consider here stock returns rescaled by the market-wide volatility for each day (see Appendix~\ref{apx:methodology}
for details). The dynamics of the market volatility is treated separately, see Appendix~\ref{sec:emp_index}. However, we have checked that the 
results below are essentially unchanged when we do not rescale single stock returns by the market wide volatility.

\subsection{The diagonal kernels}
\begin{figure}
	\center
	\includegraphics[scale=0.6]{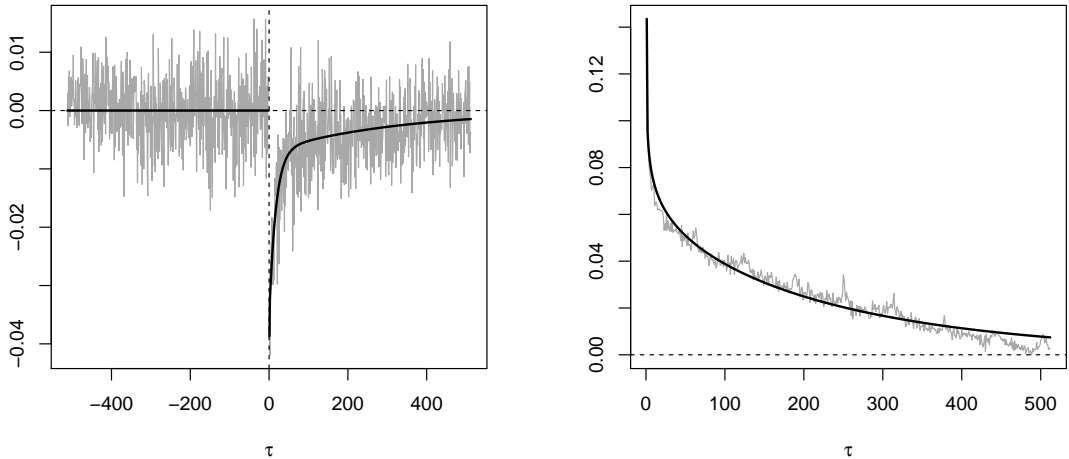}
	\caption{%**suppress the legends in the figures, notations are no longer consistent!!**
	Two empirical correlation functions: the leverage $\mathcal{L}(\tau)$ and
	         the correlation of amplitudes $\widetilde{C}^{(a)}(\tau)$, together with their fits.
	         $\mathcal{L}(\tau)$ is fitted by the sum of two exponentials $-a\,\e^{-\tau/b} -c\,\e^{-\tau/d}$, with $a=0.007$, $b=327$ days, $c=0.029$, $d=17$ days; 
	         whereas $\widetilde{C}^{(a)}(\tau)$ is 
	         fitted by a power-law truncated by an exponential: $B \tau^{-\beta}\e^{-\tau/\tau_0}$, with $\beta=0.14, B=0.106, \tau_0=290$ days.}
	\label{fig:LCa}
\end{figure}

When solving the GMM set of equations~\eqref{eq:model_predictions} in Appendix~\ref{apx:methodology}, 
we find that the diagonal elements $K(\tau,\tau)$ are an order of magnitude larger than the 
corresponding off-diagonal elements $K(\tau,\tau'\neq \tau)$. 
This was not expected {\it a priori} and is in fact one of the central result of this study. 
It confirms that {\it daily returns} indeed play a special role in the volatility feedback mechanism, as postulated by ARCH models. 
Returns on different time scales, while significant, can be treated as a perturbation of the main ARCH effect.  

This remark suggests a two-step calibration of the model: first restricting to the diagonal elements of $K$ and later including off-diagonal contributions. 
We thus neglect for a while all off-diagonal elements, and determine the $2q+1$ parameters 
$s^2, L(\tau)$ and $k(\tau)=K(\tau,\tau)$ through the following reduced GMM set of equations:
\begin{subequations}\label{eq:model_simplified}
\begin{align}\label{eq:2points_simplified}
	\vev{\sigma^2}=1&=s^2+\sum_{\tau} k(\tau) \\
	\widetilde{\mathcal{L}}(\tau)&=L(\tau) + \sum_{\tau' \neq \tau}L(\tau')\mathcal{C}^{(1)}(\tau\!-\!\tau')
	                         + \sum_{\tau'} k(\tau')\mathcal{L}(\tau\!-\!\tau')
	                         \\
	                         \label{eq:3pointsC_simplified}
	\widetilde{\mathcal{C}}^{(a)}(\tau)&\approx\sum_{\tau'}L(\tau')\mathcal{L}^{(a)}(\tau'\!-\!\tau)+
	\sum_{\tau'}k(\tau')\mathcal{C}^{(a)}(\tau-\tau').
\end{align}
\end{subequations}

\begin{figure}
	\center
	\includegraphics[scale=0.6,trim=0 0 305 0,clip]{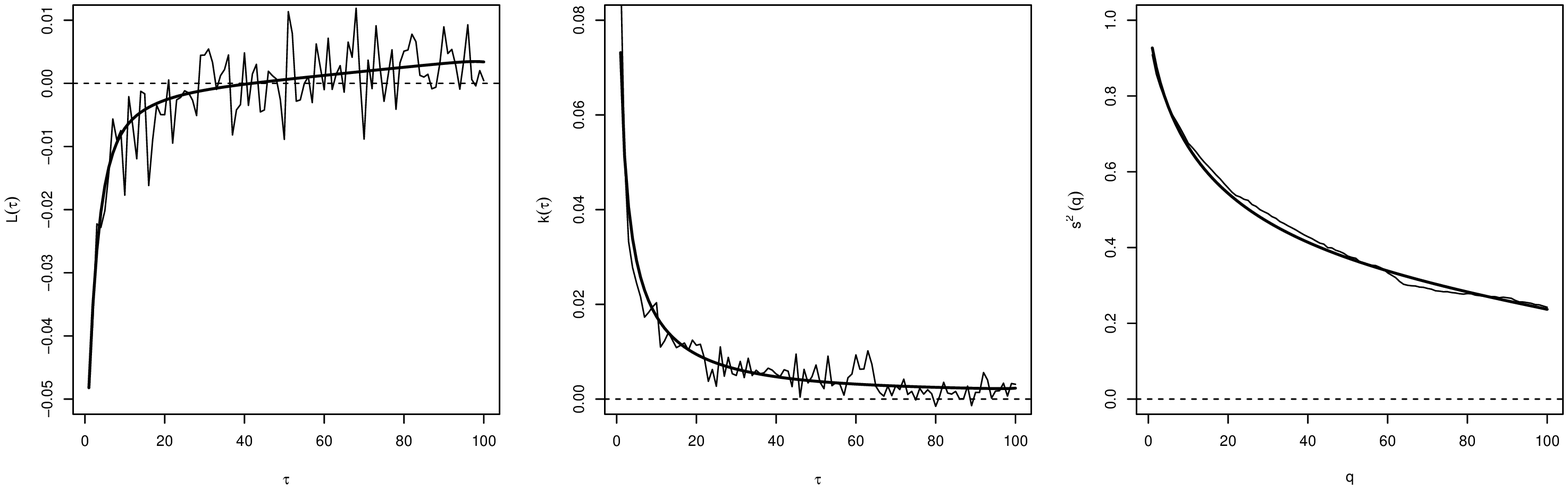}
	\caption{Calibration of the diagonal kernels for stocks, with $q=100$. 
	 \textbf{Left:} $L(\tau)$. \textbf{Right:} $K(\tau,\tau)$.
             The thick curves are obtained with fitted instead of raw input correlation functions, see Fig.~\ref{fig:LCa}.}
	\label{fig:L_stocks}
	\includegraphics[scale=0.5]{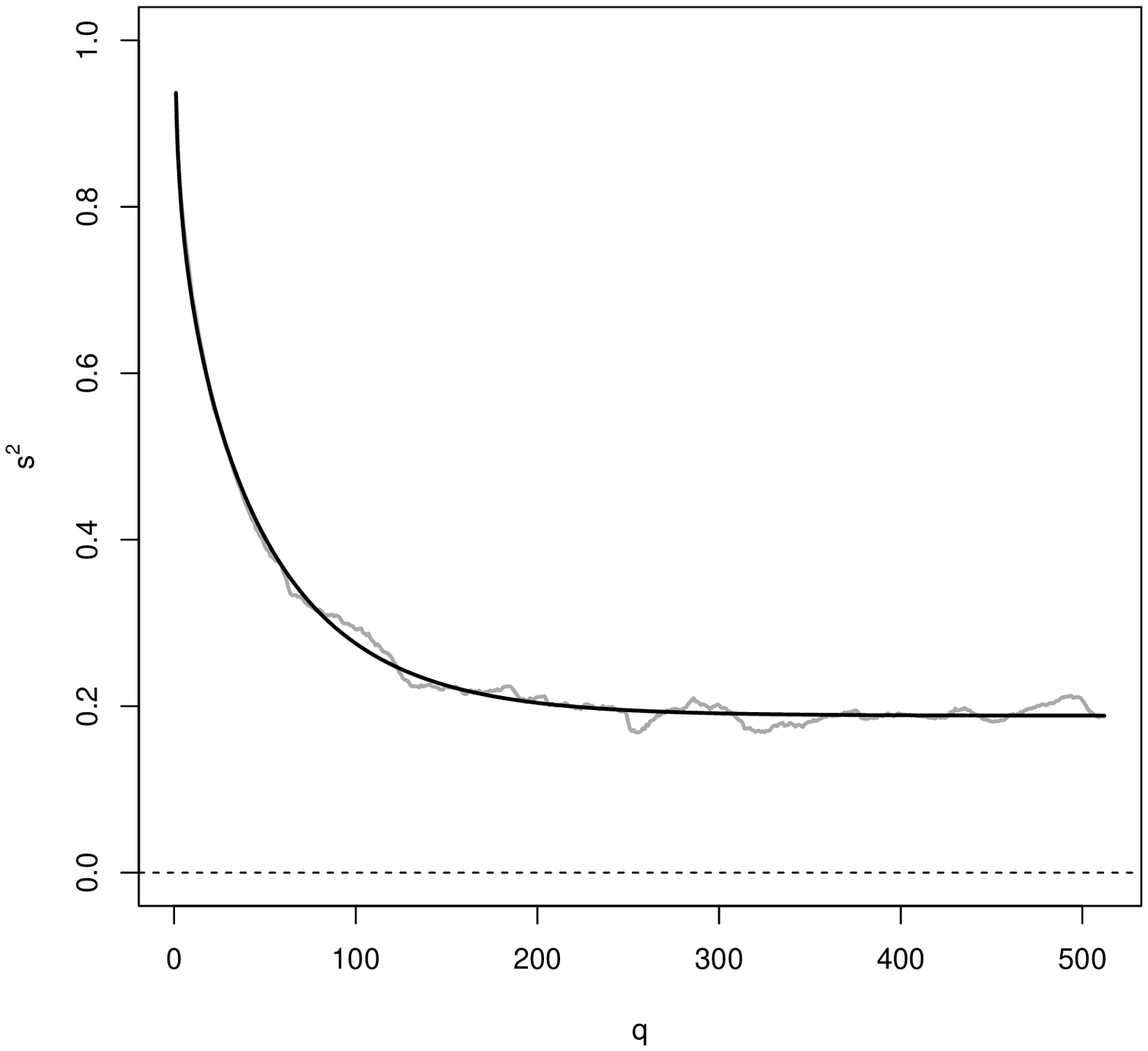}
	\caption{$s^2$ as a function of the farthest lag;
	the solid line is a fit according to the formula~\eqref{eq:fit_s2} (see text for parameter values). }
    \label{fig:s2_stocks}
\end{figure}

The input empirical correlation functions ${\mathcal{L}}(\tau)$ and $\widetilde{\mathcal{C}}^{(a)}(\tau)$ are plotted in Fig.~\ref{fig:LCa}, 
together with, respectively, a double-exponential fit and a truncated power-law fit (see legend for parameters values).
$\widetilde{\mathcal{L}}$ and $\mathcal{L}^{(a)}$ look very similar to $\mathcal{L}(\tau)$; 
note that all these functions are approximately zero for \mbox{$\tau < 0$}. 
The above equations are then solved using these analytical fits, 
which leads to the kernels $k(\tau)$ and $L(\tau)$ that we report in bold in Fig.~\ref{fig:L_stocks}. 
Using the raw data --- instead of the fits --- for all the correlation functions results in more noisy $L(\tau)$ and $k(\tau)$ (thin lines),
but still very close to the bold curves shown in Fig.~\ref{fig:L_stocks}. 
As expected, $L(\tau)$ is nearly equal to $\widetilde{\mathcal{L}}(\tau)$: 
there is a weak, but significant leverage effect for individual stocks.

\begin{figure}
 	\center
 	\includegraphics[trim=500 0 0 0,clip,scale=0.4]{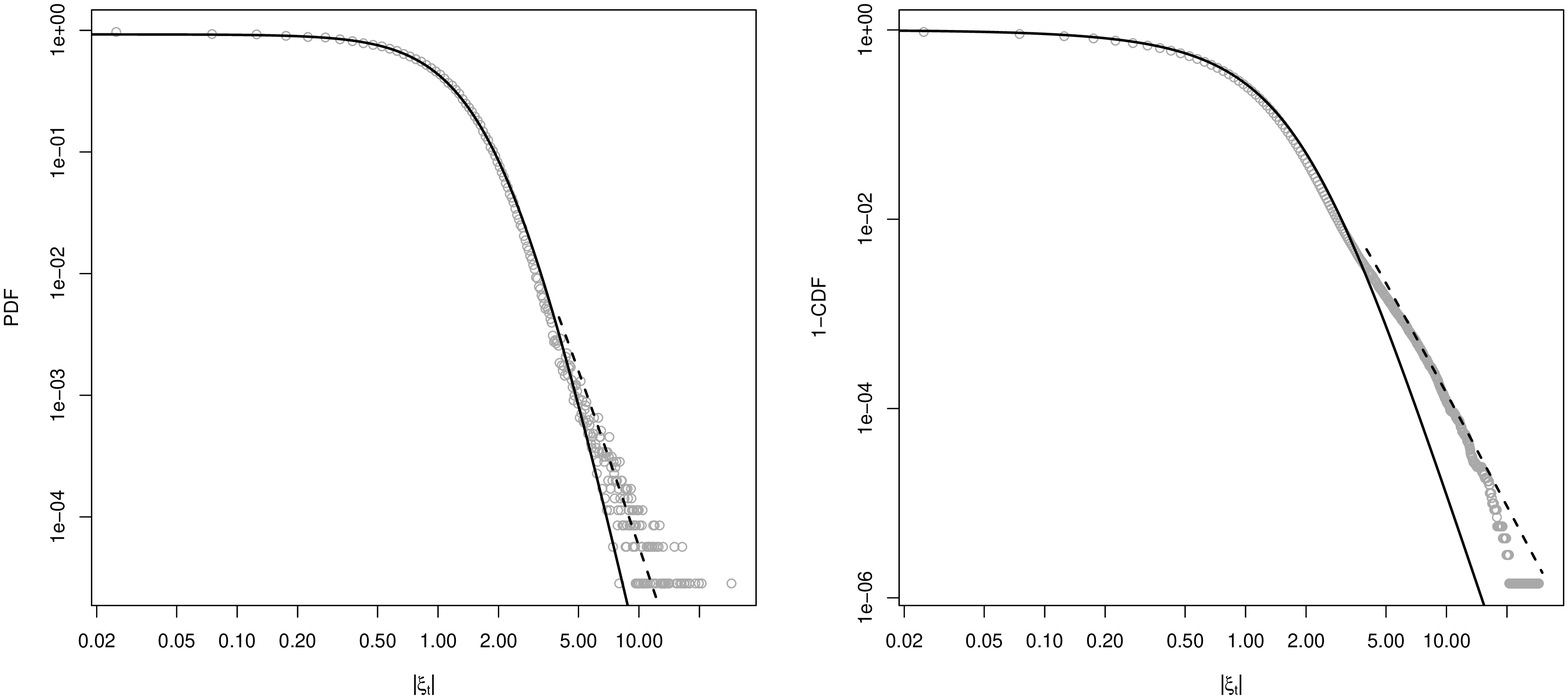}
 	\caption{Cumulative distribution function of the residuals $\xi_t=r_t/\sigma_t$. 
 	The plain line is the Student distribution with best (maximum-likelihood) fitting tail parameter $\nu=6.4$. 
 	Far tails suggest a fatter distribution with a smaller value of $\nu_\text{tail} \approx 4$ (dashed).}
 	\label{fig:mu.xi}
 \end{figure}

We then show in Fig.~\ref{fig:s2_stocks} a plot of $s^2(q) = 1 - \sum_{\tau=1}^q k(\tau)$ as a function of $q$. 
Including the feedback of the far away past  progressively decreases the value of the baseline volatility level $s^2$.
In order to extrapolate to $q \to \infty$, we have found that the following fit is very accurate:
\be\label{eq:fit_s2}
	s^2(q) = s_\infty^2 + g\,\frac{q^{1-\alpha}}{\alpha-1} \,\e^{-q/q_0},
\ee
with $s_\infty^2 \approx 0.21, \alpha \approx 1.11, g \approx 0.081$ and $q_0 \approx 53$. 
Several comments are interesting at this stage:
\begin{itemize}
\item The asymptotic value $s_\infty^2$ is equal to only $20 \%$ of the observed squared volatility%
\footnote{{When splitting daily returns into close-to-open (night) and open-to-close contributions,
we observe an even stronger feedback amplification of the overnight volatility that manifests itself by
 a nearly vanishing baseline value: most overnight price changes are news-less and purely endogenous \cite{Pierre_inprep}.
}}, 
meaning that volatility feedback effects increase the volatility by a factor \mbox{$\approx 2.25$}.  
Such a strong amplification of the volatility resonates with Shiller's ``excess volatility puzzle'' 
and gives a quantitative estimate of the role of self-reflexivity and feedback effects in financial markets 
\cite{shiller1981stock,cutler1989moves,fair2002events,soros1994alchemy,bouchaud2011endogeneous,filimonov2012quantifying,hardiman2013critical}.
\item The above fit Eq.~\eqref{eq:fit_s2} corresponds to a power-law behavior, \mbox{$k(\tau) \approx g \tau^{-\alpha}$} for $\tau \ll q_0$, 
and an exponential decay for larger lags.
Therefore, a characteristic time scale of $q_0 \approx 3$ months appears, 
beyond which volatility feedback effects decay more rapidly. 
\item With a diagonal positive kernel $K$, the condition for positive definiteness \eqref{eq:nonneg} of the quadratic form reads $\sum_{\tau}L(\tau)^2/k(\tau)\leq4s^2$.
The estimated values of $L$ and $k$ yield a left-hand side equal to 0.595, while the right-hand side amounts to 0.823. The stability criterion is therefore 
satisfied with a large margin.
\item Using the theoretical results of Appendix~\ref{sec:fourth_moment}, one can compute the theoretical value of $\langle \sigma^4 \rangle$ 
that corresponds to the empirically determined $k(\tau)$. 
As expected from the fact that $g$ is small and $\alpha$ close to unity, one finds that the fluctuations of volatility induced by the long-memory feedback are weak: 
$\langle \sigma^4 \rangle = 1.156$ (see Eq.~\ref{eq:sigma4pert} in Appendix~\ref{sec:fourth_moment}).
Including the contribution of the leverage kernel $L(\tau)$ to $\langle \sigma^4 \rangle$ does not change much the final numerical value, 
that shifts from 1.156 to 1.161. This result shows that the low-frequency (predictible) part of $\sigma^2$ has fluctuations on the order of $15 \%$. This 
may appear small, but remember that we have scaled out the market wide volatility for which the fluctuations are found to be much larger, see Appendix~\ref{sec:emp_index}.
\item The smallness of $\vev{\sigma^4}-\vev{\sigma^2}^2$ demonstrates that most of the kurtosis of the returns $r_t = \sigma_t \xi_t$ must come from the noise $\xi_t$, 
which cannot be taken as Gaussian. Using the diagonal ARCH model with the kernels determined as above to predict $\sigma_t$, 
one can study the distribution of $\xi_t = r_t/\sigma_t$ and find the most likely Student-t distribution that accounts for it. 
We find that the optimal number of degrees of freedom is \mbox{$\nu \approx 6.4$}, and the resulting fit is shown in Fig.~\ref{fig:mu.xi}. 
Note that while the body and `near-tails' of the distribution are well reproduced by the Student-t, 
the far-tails are still fatter than expected. 
This is in agreement with the commonly accepted tail index of \mbox{$\nu_\text{tail} \approx 4$}, significantly smaller than $6.4$. 
{This observation however hides a more subtle sub-daily behavior: overnight residuals are much kurtic due to rare extreme events caused by nightly news released, 
whereas open-to-close residuals are typically less fat-tailed \cite{Pierre_inprep}.}
\end{itemize}

Assuming $\xi_t$ to be a Student-t random variable with $\nu=6.4$ degrees of freedom, 
we have re-estimated $k(\tau)$ and $L(\tau)$ using maximum-likelihood (see below). 
The final results are more noisy, but close to the above ones after fitting with the same functional forms. Our  
strategy is thus to fix both $k(\tau)$ and $L(\tau)$ to their above values, and only focus on the {\it off-diagonal} elements of $K$ henceforth. 

\subsection{The off-diagonal kernel, GMM \& maximum likelihood estimation}\label{ssec:GMMML}

We can now go back to Eq.~\eqref{eq:4points} that allows one to solve for $K(\tau',\tau'' > \tau')$, once $k(\tau)$ and $L(\tau)$ are 
known. As announced above, we choose $q=20$ for the time being. Because $\mathcal{D}$ involves the fourth moment of the returns, this GMM procedure
is quite unstable, even with a lot of data pooled together, and even after the truncation of large returns. Maximum likelihood estimates (that puts less
weight on extreme events) would be more adapted here, but the dimensionality of the problem prevents a brute force determination of the $q(q-1)/2$ parameters.

In order to gain some robustness, we use the following strategy. The Student log-likelihood per point ${\mathcal{I}}$, with $\nu$ degrees of freedom, is given by:\footnote{In the following
we do {\it not} truncate the large returns, but completely neglect the weak linear correlations $\mathcal{C}^{(1)}(\tau)$ that are present for small lags, and
that should in principle be taken into account in the likelihood estimator.}
\be\label{eq:loglike}
{\mathcal{I}}_\nu(L,\mathcal{K},\{r_t\})=\frac{1}{2T}\sum_t\left[\nu\,\ln{a_t^2}-(\nu+1)\,\ln(a_t^2+r_t^2)\right],\qquad a_t^2 \equiv(\nu-2)\,\sigma_t^2,
\ee
where $r_t\equiv \sigma_t\xi_t$ and $\sigma_t^2$ is given by the QARCH model expression, Eq.~\eqref{eq:quadraticARCH}, 
and $\mathcal{K}$ is a notation for the off-diagonal content only. 
We fix $\nu=6.4$ and determine numerically the gradient 
$\partial {\mathcal{I}}_\nu$
and the Hessian $\partial \partial {\mathcal{I}}_\nu$ of ${\mathcal{I}}_\nu$ as a function of {\it all} the $q(q-1)/2$ off-diagonal coefficients 
$K(\tau',\tau'' > \tau')$, computed either around the GMM estimates of these parameters, or around the ARCH point where all these coefficients are zero. 
Note that $\partial {\mathcal{I}}_\nu$ is a vector with $q(q-1)/2$ entries and $\partial \partial {\mathcal{I}}_\nu$ is a $q(q-1)/2 \times q(q-1)/2$
matrix. It so happens that the eigenvalues of the Hessian are all found to be {\it negative}, i.e.\ the starting point is in the vicinity of a local
maximum. This allows one to find easily the values of the $q(q-1)/2$ parameters that maximize the value of ${\mathcal{I}}_\nu$; they are (symbolically) given by:
\be
\mathcal{K}^* = \mathcal{K}_0 - \left(\,\overline{\partial \partial {\mathcal{I}}_\nu}\,\right)^{-1} \cdot\overline{ \partial {\mathcal{I}}_\nu},
\ee
where $\mathcal{K}_0$ is the chosen starting point --- either the GMM estimate $\mathcal{K}_0=\mathcal{K}_{\text{GMM}}$ based on Eq.~\eqref{eq:4points}, 
or simply $\mathcal{K}_0=0$ if one starts from a diagonal ARCH model --- and the overline on top of $\partial\partial\mathcal{I},\partial\mathcal{I}$ indicates averaging over stocks. 
This one step procedure is only approximate but can be iterated; 
it however assumes that the maximum is close to the chosen initial point, 
and would not work if some eigenvalues of the Hessian become positive. 
In our case, both starting points ($\mathcal{K}_0=0$ or $\mathcal{K}_0=\mathcal{K}_{\text{GMM}}$) lead to nearly exactly the same solution; 
furthermore the Hessian recomputed at the solution point is very close to the initial Hessian, 
indicating that the likelihood is a locally quadratic function of the parameters, 
and the gradient evaluated at the solution point is very close to zero in all directions,
confirming that a local maximum as been reached.\footnote{A discussion of the bias and error of the MLE-estimated parameters is given in Appendix~\ref{ssec:errorMLE}.}

The most likely off-diagonal coefficients of ${K}^*$ are found to be highly significant (see Tab.~\ref{tab:IS-OOS}): 
the IS increase of likelihood from the purely diagonal ARCH($q$) model is $\Delta \mathcal{I} \approx 10^{-3}$ per point. 
This is confirmed by an out-of-sample (OS) experiment, where we determine $\mathcal{K}^*$ on half the pool of stocks 
and use it to predict the volatility on the other half.% (whence the above factor ${1}/{2}$ in the numerical estimation of number $n$). 
 The experiment is performed over $N_{\text{samp}}=150$ random pool samplings. 
The average OS likelihood is very significantly better for the full off-diagonal kernel $\mathcal{K}^*$ than for the diagonal ARCH($q$), 
itself being better than the GMM estimate $\mathcal{K}_{\text{GMM}}$ based on Eq.~\eqref{eq:4points}, and probably subject to biases due to the
truncation procedure. Note that the full off-diagonal kernel $\mathcal{K}^*$ has many more parameters than the diagonal ARCH($q$); 
it therefore starts with a handicap out-of-sample because of the bias on the OS likelihood being proportional to $M$, see Appendix~\ref{ssec:errorMLE}.

\begin{figure}
 	\center
 	\includegraphics*[scale=0.35,angle=-90]{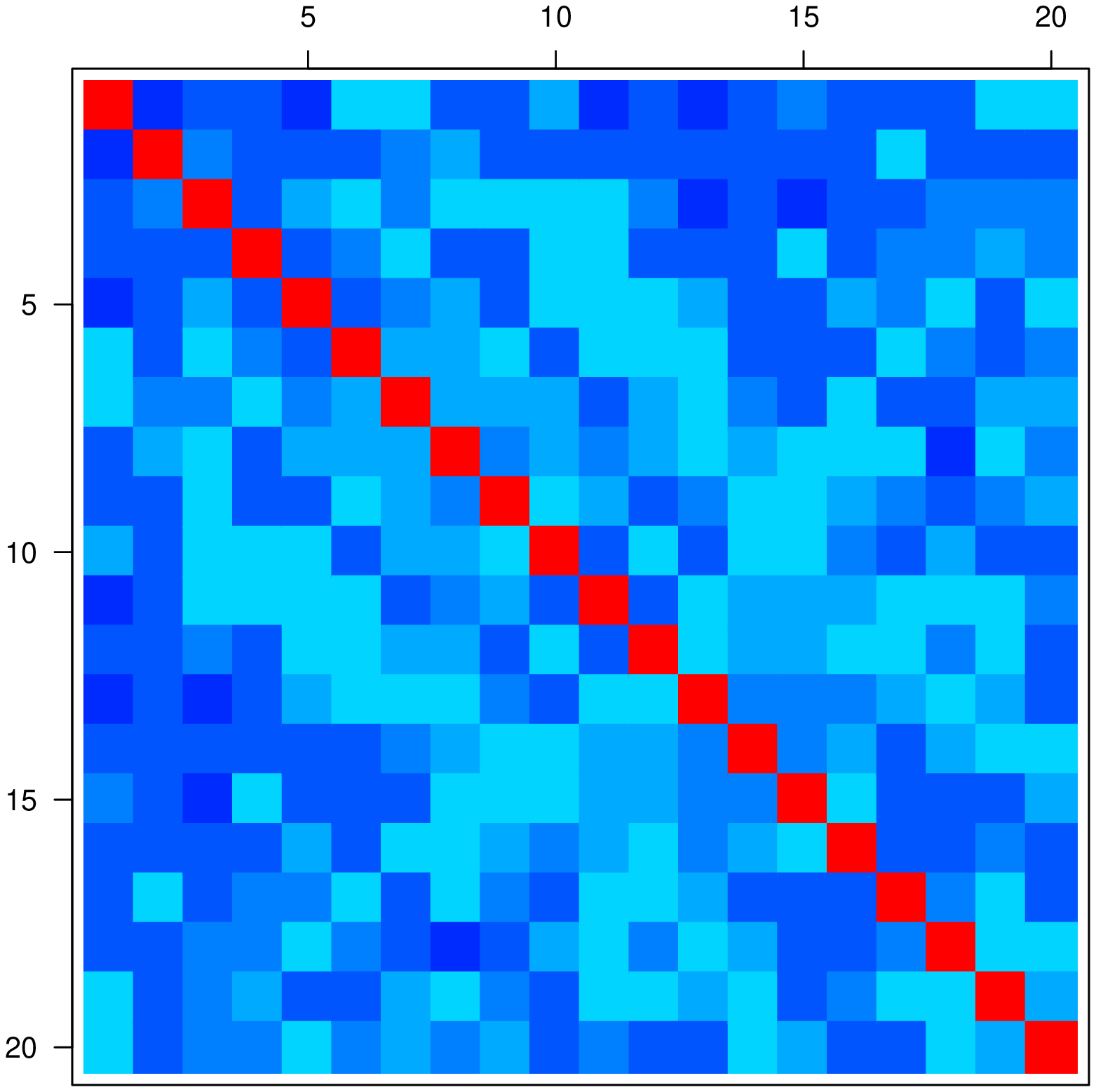}
 	\includegraphics*[scale=0.35,angle=-90]{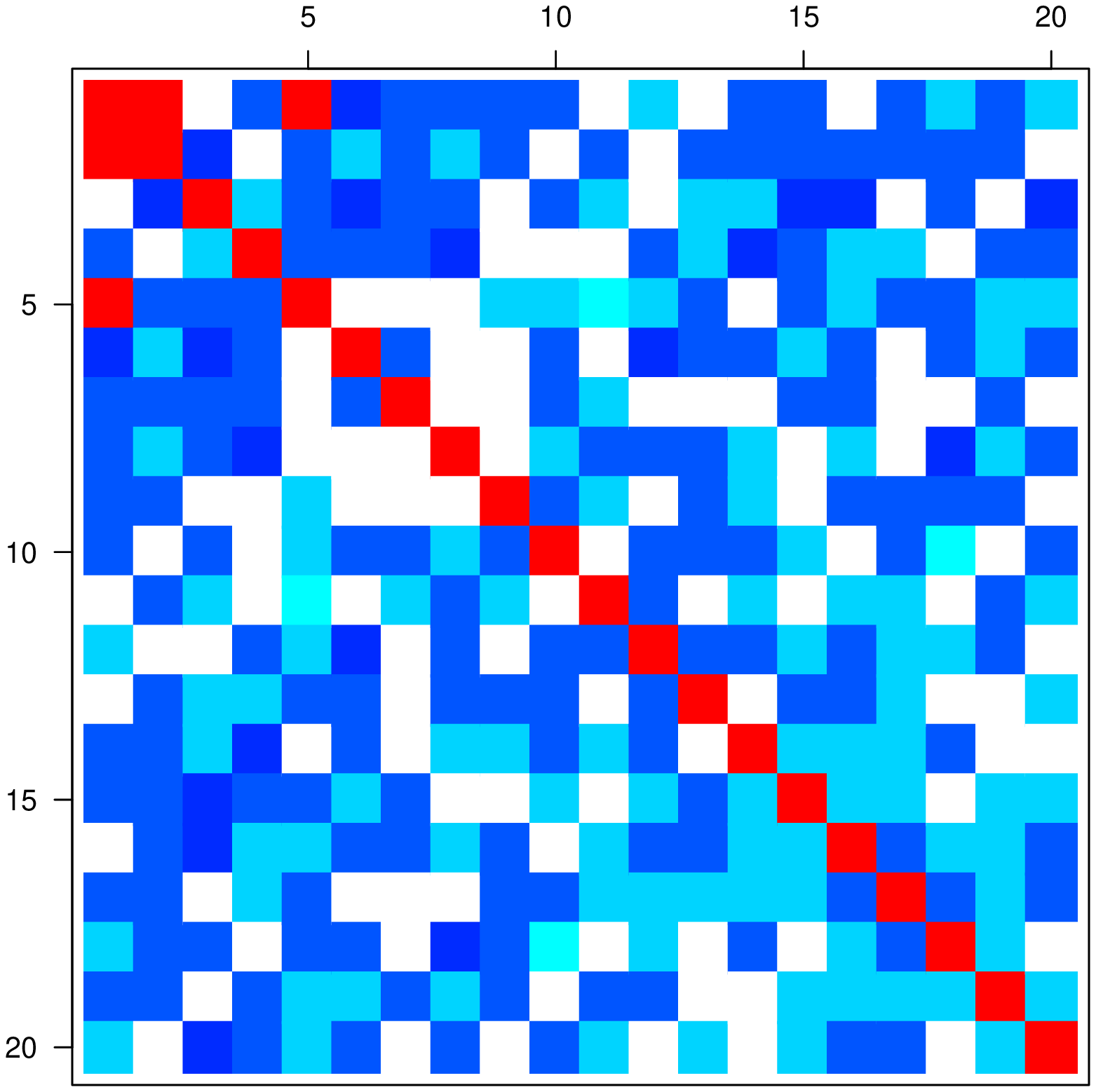}
 	\includegraphics*[scale=0.35,angle=-90]{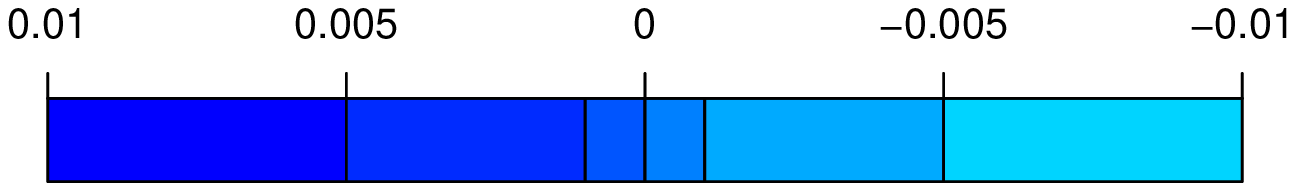}
 	\caption{Heatmap of the unconstrained model.
 	\textbf{Left: } GMM estimation; \textbf{Right: } ML estimation (white spots correspond to values smaller than their corresponding error margin). Red spots correspond to values 
 	larger than $0.01$. Note the 
 	negative streaks at large lags, and the significant off-diagonal entry for $\tau=1,\tau'=2$ and $\tau=1,\tau'=5$. }
 	\label{fig:heatmap}
 \end{figure}

However, as announced above, the off-diagonal elements of ${K}^*$ are a factor ten smaller than the corresponding diagonal values.
We give a heat-map representation of the matrix ${K}^*$ in Fig.~\ref{fig:heatmap}. Various features are immediately apparent. 

First,the most significant off-diagonal terms correspond to $K(1,2)=K(2,1)$ and $K(1,5)=K(5,1)$, showing that the day before last, and the same day a week before ($5$ trading days) play a special role. 

Second, while the off-diagonal elements are mostly positive for small $\tau',\tau''$, 
clear negative streaks appear for intermediate and large $\tau$s. 
This is unexpected, since one would have naively guessed that any trend 
(i.e.\ positive realized correlations between returns) should {\it increase} future volatilities. 
Here we see that some quadratic combinations of past returns contribute negatively to the volatility. 
This will show up in the spectral properties of $K^*$ (see Section~\ref{sec:spectral2} below).

\begin{figure}
	\center
	\includegraphics[scale=0.45]{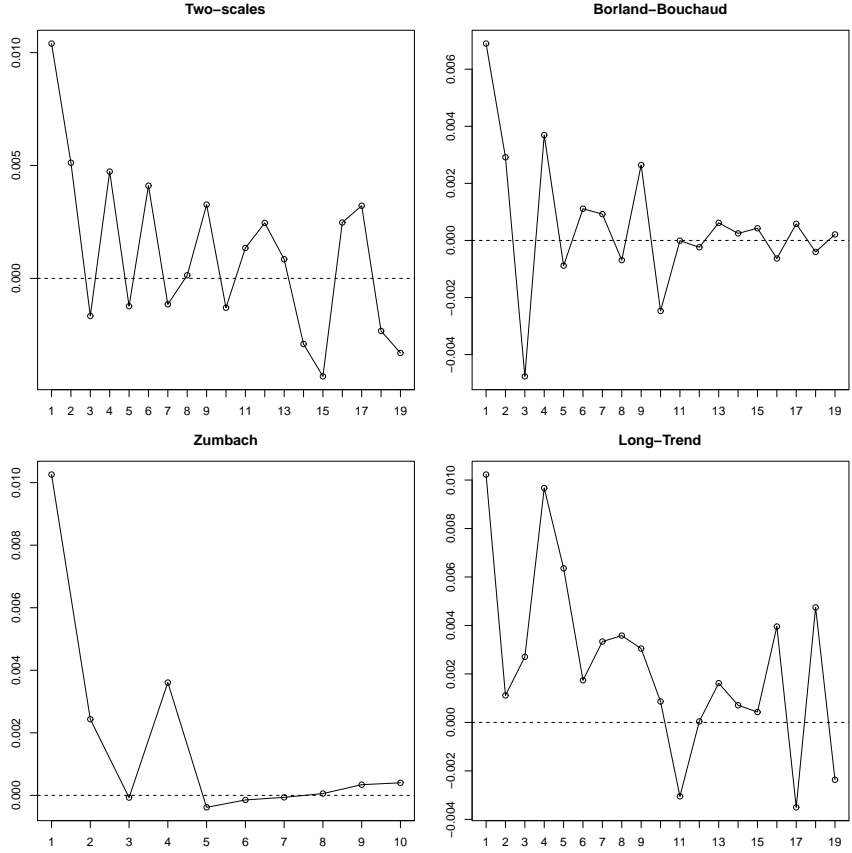}
	\caption{Plot of the empirically determined kernels $g_2(\ell)$, $g_\text{BB}(\ell)$, $g_\text{Z}(\ell)$ and $g_\text{LT}(\ell)$ for the 
	restricted models of Section~\ref{sec:section3}.}
	\label{fig:structure_fcts}
\end{figure}

The third surprise is that there does not seem to be any obvious structure of the matrix, 
that would be reminiscent of one of the simple models represented in Fig.~\ref{fig:matrices}.
This means that the fine structure of volatility feedback effects is much more subtle than anticipated. This conclusion is unchanged 
when daily returns are not rescaled by the market-wide volatility on the same day.
We have nevertheless implemented a {\it restricted} maximum-likelihood estimation 
that imposes the structure of one of the models considered in Section~\ref{sec:section3}. 
We find that all these models are equally ``bad'' --- 
although they lead to a significant increase of likelihood compared to the pure ARCH case, both IS and OS, 
they are all superseded by the unconstrained model shown in Fig.~\ref{fig:heatmap}, again both IS and OS (see Tab.~\ref{tab:IS-OOS}). 
The best OS model (taken into account the number of parameters) is ``Long-trend'', with a kernel $g_\text{LT}(\ell)$ shown in Fig.~\ref{fig:structure_fcts}, 
together with the functions $g_2(\ell)$, $g_\text{BB}(\ell)$, $g_\text{Z}(\ell)$. 
While $g_\text{LT}(\ell)$ looks roughly like an exponential with memory time $10$ days, the two-day return kernel $g_2(\ell)$ reveals intriguing oscillations. 
Two-day returns influence the volatility quite differently from one day returns!
On the other hand, we do not find any convincing sign of the multiscale ``BB'' structure postulated in \cite{borland2005multi}.\footnote{In fact, when daily returns are
not rescaled by the market-wide volatility on the same day, the BB model becomes marginally the best one Out-of-Sample, but only very marginally.}
Note that the structure shown in Fig.~\ref{fig:structure_fcts} is found to be stable when $q$ is changed. 
It would be interesting to subdivide the pool of stocks in different categories 
(for example, small caps/large caps) or in different sub-periods, and study how the off-diagonal structure of $K$ is affected.
However, we note that the dispersion of likelihood over different samplings of the pool of stocks is only $50\%$ larger than 
the ``true'' dispersion, due only to random samplings of a fixed QARCH model with parameters calibrated to the data (see caption of Tab.~\ref{tab:IS-OOS}). 
This validates, at least as a first approximation, the assumption of homogeneity among all the stock series that have been averaged over.

%Finally, we have revisited the most likely value of the Student parameter $\nu$ with now the full matrix $K^*$, plus diagonal terms up to 
%$q=100$, and found again $\nu=6.4$. This shows that our procedure is consistent from that point of view.

To conclude this empirical part, we have performed several ex-post checks to be sure that our assumptions and preliminar estimations are justified.
First, we have revisited the most likely value of the Student parameter $\nu$ 
(tail index of the distribution of the residuals $\xi(t)=r(t)/\sigma_{\text{QARCH}}(t)$)
with now the full matrix $K^*$, plus diagonal terms up to $q=100$, and found again $\nu=6.4$. 
This shows that our procedure is consistent from that point of view.
Second, we have computed the quadratic correlation of the residuals $\xi_t$, which are assumed in the model to be IID random variables with, in 
particular, no variance autocorrelation: $\vev{\xi_t^2\xi_{t-\tau}^2}-1=0$ for $\tau \neq 0$. 
Empirically, we observe a negative correlation of weak magnitude exponentially decaying with time. 
This additional dependence of the amplitude of the residuals, together with the excess fat tails of their probability distribution,
is probably a manifestation of the truly exogenous events occuring in financial markets that have different statistical properties \cite{joulin2008stock} 
and not captured by the endogeneous feedback mechanism. Finally, about the universality hypothesis, we discuss in the caption of Tab.~\ref{tab:IS-OOS} how the assumption of homogeneous stocks
is justified by comparing the cross-sectional dispersion of the likelihoods obtained empirically and on surrogate simulated series.

\begin{table}
\begin{center}
\begin{tabular}{c||cc|c|}
  &GMM        &ARCH(20)    &ARCH+ML\\\hline\hline
IS&$-1.31533$ &$-1.31503$ &$\mathbf{-1.31405}$\\
OS&$-1.32003$ &$-1.31971$ &$\mathbf{-1.31914}$\\\hline
\end{tabular}

\vspace{2em}

\begin{tabular}{c|cccccc|}
   &ARCH+BB            &ARCH+Z     &ARCH+2s    &ARCH+LT     &ARCH+2s+Z    &ARCH+2s+LT\\\hline\hline
IS&$        -1.31486 $ &$-1.31490$ &$-1.31490$ &$-1.31487$  &$-1.31488$   &$-1.31489$\\
OS&$       {-1.31960}$ &$-1.31957$ &$-1.31962$ &$-1.31957$  &$-1.31956$   &$-1.31957$\\\hline
\end{tabular}
\caption{Log-likelihoods, according to Eq.~\eqref{eq:loglike}. 
In-sample and out-of-sample likelihoods are computed as follows: 
for each of $N_{\text{samp}}=150$ iterations, half of the stock names are randomly chosen for the calibration of $\mathcal{K},L$ 
and the likelihood is computed with the obtained $\mathcal{K}^*,L^*$ on each series of the same sample (`In-sample' likelihoods).
Then, the likelihood is again computed with the same parameters but on the series of the other sample (`Out-of-sample' likelihoods).
While the former quantify how much the estimated model succeeds in reproducing the given sample, 
the latter measure the reliability of the model on \emph{other} similar datasets. 
In order to quantify the validity of the model in an absolute way, the likelihood can be compared with the ``true'' value,  
obtained with simulated data (since an analytical treatment is out of reach). 
The average likelihood per point $\overline{\mathcal{I}}^*(r_t)$ with $r_t$ simulated as a QARCH with parameters $K^*,L^*$, 
and $\nu=6.4$ is equal to $-1.34019$, which is 1.5\% away from the empirical values\protect\footnotemark.
The likelihoods reported in the table are averages over all samplings, and the corresponding 1-s.d.\ dispersion is found to be $\approx 3\cdot 10^{-3}$
in all cases, to be compared to  $2\cdot 10^{-3}$ for random surrogate samplings of a fixed QARCH model with the same parameters.
}
\label{tab:IS-OOS}
\end{center}
\end{table}
\footnotetext{Note that the true likelihood is not necessarily larger than the realized one under a misspecified model.}

\subsection{Spectral properties of the empirical kernel $K$}
\label{sec:spectral2}
\begin{figure}
	\center
	\includegraphics[scale=0.4]{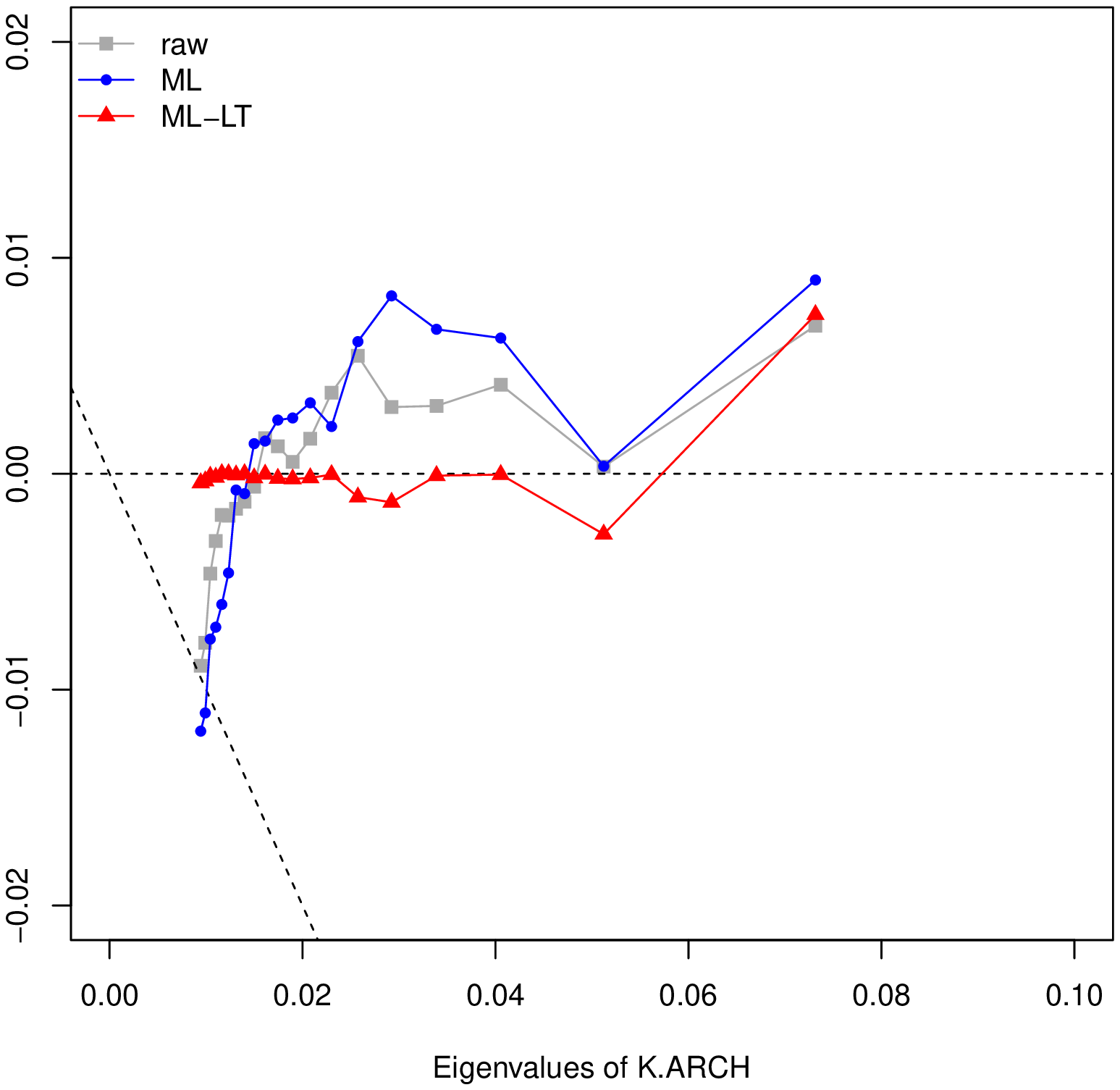}
	\includegraphics[scale=0.3]{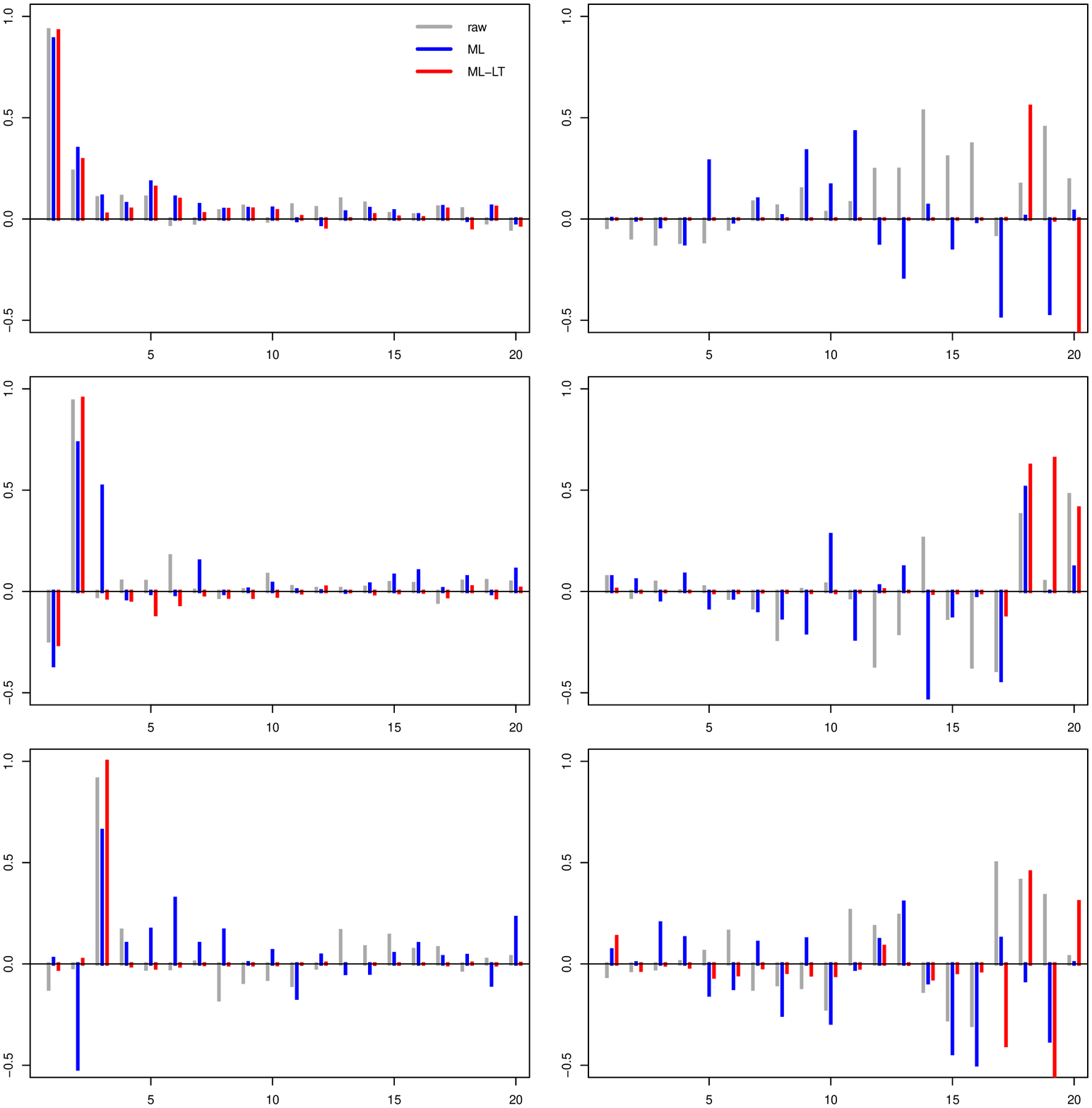}
	\caption{Spectral decomposition of the feedback kernel $K$, for the GMM, ML and ML+LT estimates. 
	\textbf{Left: }  The difference between the ranked eigenvalues of the estimated kernels and those of $K_{\text{ARCH}}$ 
	                 as a function of the latter (again ranked).
	                 The dashed oblique line has slope $-1$ and separates positive eigenvalues from negative ones.
	\textbf{Right: } Structure of the first three and last three eigenvectors. 
	                 Whatever the estimation method, the first eigenvector has a non trivial structure, with mostly positive components, indicating 
	a genuine departure from the diagonal ARCH benchmark, for which we would find a single peak at $\tau=1$.}
	\label{fig:eigen_stocks}
\end{figure}

As discussed in Section~\ref{sec:spectral}, another way to decipher the structure of $K$ is to look at its eigenvalues and eigenvectors. 
We show in Fig.~\ref{fig:eigen_stocks} the eigenvalues of $K^*$ as a function of the eigenvalues of the purely diagonal ARCH model. 
We see that a) the largest eigenvalue is clearly shifted upwards by the off-diagonal elements; 
            b) the structure of the top eigenvector is non-trivial, and has positive contributions at all lags (up to noise);
            c) the unconstrained estimations --- both GMM and ML --- lead to 6 very small eigenvalues (perhaps even slightly negative for ML) 
               that all constrained models fail to reproduce.
               
The positiveness of all eigenvalues was far from granted a priori, 
because nothing in the calibration procedure imposes the positivity of the matrix $K^*$. 
Although we naively expected that past excitations could only lead to an amplification of future volatility
(i.e.\ that only strictly positive modes should appear in the feedback kernel), we observe that quasi-neutral modes do occur and
appear to be significant. 
This is clearly related to the negative streaks noted above at large lags, 
but we have no intuitive interpretation for this effect at this stage.

\section{Time-reversal invariance}
\label{sect:TRI_0}

\begin{figure}
	\center
	\includegraphics[scale=0.4]{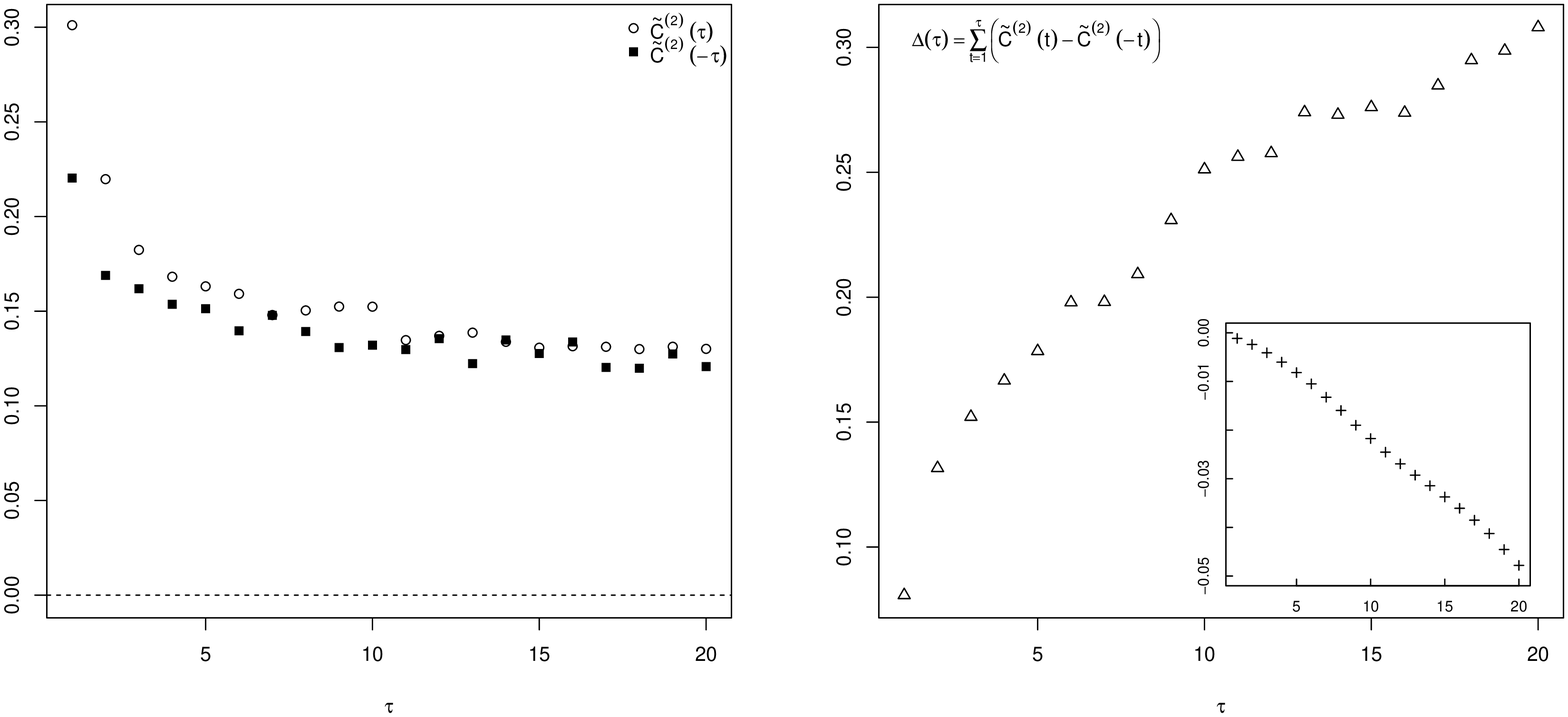}
	\caption{Measure of time-reversal asymmetry (Eq.~\ref{eq:tra}) for the stock data.
	         \textbf{Inset: } The contribution to $\Delta(\tau)$ stemming from the leverage, i.e.\ the quantity 
	         $\sum\limits_{\tau'=1}^\tau L(\tau')\left[\mathcal{L}(\tau'\!-\!\tau)- \mathcal{L}(\tau'\!+\!\tau)\right]$.
	         Note that this contribution is negative, and an order of magnitude smaller than $\Delta(\tau)$ itself.}
	\label{fig:TRI_emp}
%\end{figure}
%\begin{figure}
%	\center
	\includegraphics[scale=0.4]{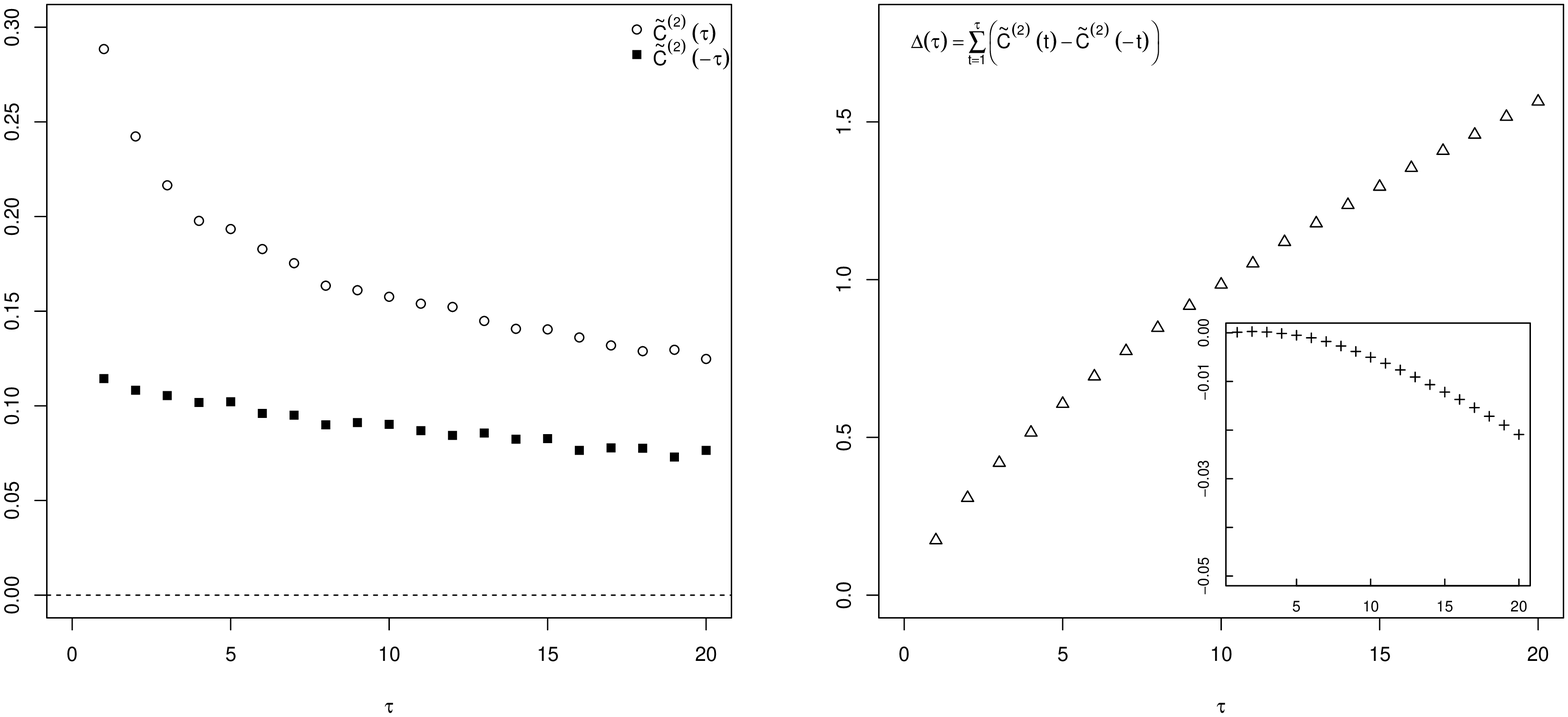}
	\caption{Measure of time-reversal asymmetry (Eq.~\ref{eq:tra}) for a simulated ARCH model with Student ($\nu=4$) residuals on the 5 minute scale.
	The parameters of the simulation are the estimated kernel $k^*(\tau)$ and $L^*(\tau)$ for stocks, with $q=20$.
	At each date, 100 intraday prices are simulated (corresponding to the number of 5-minutes bins inside 8 hours) with the {\it same} $\sigma_t^2$ 
	given by the QARCH model. The volatility is then computed using Rogers-Satchell's estimator, exactly as for empirical data.}
	\label{fig:TRI}
\end{figure}

As noticed in the introduction, QARCH models violate, by construction, time-reversal invariance. 
Still, the correlation of the squared returns ${\mathcal{C}}^{(2)}(\tau)$ is trivially invariant 
under time-reversal, i.e.\  ${\mathcal{C}}^{(2)}(\tau)={\mathcal{C}}^{(2)}(-\tau)$. 
However, the correlation of the true squared volatility with past squared returns $\widetilde{\mathcal{C}}^{(2)}(\tau)$ 
is in general not (see \cite{pomeau1985symetrie,zumbach2009time} for a general discussion). 
A measure of TRI violations is therefore provided by the integrated difference $\Delta(\tau)$:
\be\label{eq:tra}
\Delta(\tau) = \sum_{\tau'=1}^\tau \left[\widetilde{\mathcal{C}}^{(2)}(\tau')- \widetilde{\mathcal{C}}^{(2)}(-\tau')\right].
\ee
The empirical determination of $\widetilde{\mathcal{C}}^{(2)}(\tau)$ and $\Delta(\tau)$ for stock returns is shown in Fig.~\ref{fig:TRI_emp}. 
Although less strong than for simulated data (see Fig.~\ref{fig:TRI}), we indeed find a clear signal of TRI violation for stock returns, 
in agreement with a related study by Zumbach~\cite{zumbach2009time}.
We compare in Fig.~\ref{fig:TRI} the quantity $\Delta(\tau)$ obtained from a {\it bona fide} numerical simulation of the model, 
with previously estimated parameters. 
Note that any measurement noise on the volatility $\sigma_t^2$ tends to reduce the TRI violations, 
but we have performed the numerical simulation in a way to reproduce this measurement noise as faithfully as possible.

However, the alert reader should worry that the existence of asymmetric leverage correlations ${\mathcal{L}}(\tau >0) \neq 0$ 
between past returns and future volatilities is in itself a TRI-violating mechanism, which has nothing to do with the ARCH feedback mechanism. 
In order to ascertain that the effect we observe is not a spurious consequence of the leverage effect, 
we have also computed the contribution of ${\mathcal{L}}(\tau)$ to $\Delta(\tau)$, which reads to lowest order and schematically:
\be
\Delta(\tau) = \sum\limits_{\tau'=1}^\tau L(\tau')\left[\mathcal{L}(\tau'\!-\!\tau)- \mathcal{L}(\tau'\!+\!\tau)\right] + K \,\,\text{contributions}.
\ee
The first term on the right-hand side is plotted in the inset of Fig.~\ref{fig:TRI_emp}, 
and is found to have a {\it negative} sign, and an amplitude much smaller than $\Delta(\tau)$ itself. 
It is therefore quite clear that the TRI-violation reported here is genuinely associated to the ARCH mechanism and not to the leverage effect, 
a conclusion that concurs with that of \cite{zumbach2009time}.

Still, the smallness of the empirical asymmetry compared with the simulation results suggests that the ARCH mechanism is ``too deterministic''.
It indeed seems reasonable to think that the baseline volatility $s^2$ has no reason to be constant, but may contain an extra random contribution.
Writing
\[
	\sigma^2(t)=\sigma_{\text{A}}^2(t) + \omega(t); \qquad \vev{\omega}=0; \qquad\vev{\omega(t)\omega(t-\tau)}\equiv\mathcal{C}_{\omega}(\tau)=\mathcal{C}_{\omega}(-\tau)
\]
with $\omega_t$ a noise contribution and $\sigma_{\text{A}}$ the ARCH volatility\footnote{
For the sake of clarity we consider here a diagonal ARCH framework, but the argument is straightforwardly generalized for a complete QARCH.}
(i.e.\ deterministic when conditioned on past returns), then the asymmetry is found to be given by:
\[
	\Delta(\tau)=\Delta_{\text{A}}(\tau)-\sum_{\tau''=1}^\tau\sum_{\tau'=1}^qk(\tau')\left[\mathcal{C}_{\omega}(\tau'-\tau'')-\mathcal{C}_{\omega}(\tau'+\tau'')\right].
\]
If one assumes that the correlation function $\mathcal{C}_{\omega}$ is positive and decays with time, 
the extra contribution to the asymmetry is negative, and reduces the observed TRI.
This conclusion speaks in favor of a mixed approach to volatility modeling, 
bringing together elements of autoregressive QARCH models with those of stochastic volatility models.
It would in fact be quite surprising that (although unobservable) the volatility should be a purely deterministic function of past returns.
Although the behavioral interpretation of the above construction is not clear at this stage, 
the uncertainty on the baseline volatility level $s^2$ could come, for example, 
from true exogenous factors that mix in with the volatility feedback component described by the QARCH framework.

\section{Conclusion, extensions}

We have revisited the QARCH model, which postulates that the volatility today can be expressed as a general quadratic form of the past daily returns $r_t$. 
The standard ARCH or GARCH framework is recovered when the quadratic kernel is diagonal, 
which means that only past squared daily returns feedback on today's volatility. 
This is a very restrictive {\it a priori} assumption, and the aim of the present study was
to unveil the possible influence of other quadratic combinations of past returns, such as, for example, square weekly returns. 
We have defined and studied several sub-families of QARCH models that make intuitive sense, 
and with a restricted number of free parameters. 

The calibration of these models on US stock returns has revealed several features that we did not anticipate. 
First, although the off-diagonal (non ARCH) coefficients of the quadratic kernel are found to be highly significant 
both in-sample and out-of-sample, they are one order of magnitude smaller than the diagonal elements. 
This confirms that daily returns indeed play a special role in the volatility feedback mechanism, as postulated by ARCH models. 
Returns on different time scales can be thought of as a perturbation of the main ARCH effect. 
The second surprise is that the structure of the quadratic kernel does not abide with any of the simpler QARCH models proposed so-far in the literature. 
The fine structure of volatility feedback is much more subtle than anticipated. 
In particular, neither the model proposed in \cite{borland2005multi} (where returns over several horizons play a special role), 
nor the trend model of Zumbach  in \cite{zumbach2010volatility} are supported by the data. 
The third surprise is that some off-diagonal coefficients of the kernel are found to be negative for large lags, 
meaning that some quadratic combinations of past returns contribute negatively to the volatility. 
This also shows up in the spectral properties of the kernel, which is found to have very small eigenvalues, 
suggesting the existence of unexpected volatility-neutral patterns of past returns.  

As for the diagonal part of the quadratic kernel, our results are fully in line with all past studies: 
the influence of the past squared-return $r_{t-\tau}^2$ on today's volatility $\sigma_t^2$ 
decays as a power-law $g\, \tau^{-\alpha}$ with the lag $\tau$, at least up to $\tau \approx 2$ months, 
with an exponent $\alpha$ close to unity ($\alpha \approx 1.11$), 
which is the critical value below which the volatility diverges and the model becomes non-stationary. 
As emphasized in \cite{borland2005multi}, markets seem to operate close to criticality 
(this was also noted in different contexts, see \cite{bouchaud2004fluctuations,bouchaud2006random, bacry2012non,toth2011anomalous, parisi2012universality} for example, 
and more recently \cite{hardiman2013critical}). 
The smallness of $\alpha - 1$ has several important consequences: first, this leads to long-memory in the volatility; 
second, the average square volatility is a factor 5 higher than the baseline volatility, 
in line with the excess volatility story \cite{shiller1981stock}: 
most of the market volatility appears to be endogenous and comes from self-reflexive, 
feedback effects (see e.g.\ \cite{soros1994alchemy,bouchaud2011endogeneous,filimonov2012quantifying,hardiman2013critical} and references therein). 
Third, somewhat paradoxically, the long memory nature of the kernel leads to \emph{small} fluctuations of the volatility. 
This is due to a self-averaging mechanism occurring in the feedback sum, that kills fat tails.
This means that the high kurtosis of the returns in ARCH models cannot be ascribed to volatility fluctuations 
but rather to leptokurtic residuals, also known as unexpected price jumps. 

Related to price jumps, we should add the following interesting remark that stresses the difference 
between endogenous jumps and exogenous jumps within the ARCH framework. 
Several studies have revealed that the volatility relaxes after a jump as a power-law, 
akin to Omori's law for earthquake aftershocks: 
$\sigma^2_{\tau}\sim \sigma^2_{0}\tau^{-\theta}$, where $t=0$ is the time of the jump. 
The value of the exponent $\theta$ seems to depend on the nature of the initial price jump. 
When the jump occurs because of an exogenous news, $\theta \approx 1$ \cite{lillo2003power,joulin2008stock}, 
whereas when the jump appears to be of endogenous origin, the value of $\theta$ falls around 
$\theta \approx \frac12$ \cite{zawadowski2006short,joulin2008stock}. 
In other words, as noted in \cite{joulin2008stock}, the volatility seems to resume its normal course faster after a news 
than when the jump seems to come from nowhere. 
A similar difference in the response to exogenous and endogenous shocks was also reported in \cite{sornette2004endogenous} for book sales. 
Now, if one simulates long histories of prices generated using an ARCH model with a diagonal kernel decaying as $g\, \tau^{-\alpha}$, 
one can measure the exponent $\theta$ by conditioning on a large price jump (which can only be endogenous, by definition!). 
One finds that $\theta$ varies continuously with the amplitude of the initial jump, 
and saturates around $\theta \approx \frac12$ for large jumps (we have not found a way to show this analytically). 
A similar behavior is found within multifractal models as well \cite{sornette2003what}. 
If on the other hand an exogenous jump is artificially introduced in the time series by imposing a very large value of $\xi_{t=0}$, 
one expects the volatility to follow the decay of the kernel and decay as $g\,\tau^{-\alpha} \xi_0^2$, leading to $\theta=\alpha \approx 1$. 
Therefore, the dichotomy between endogenous and exogenous shocks seem to be well reproduced within the ARCH framework. 

Finally, we have emphasized the fact that QARCH models are by construction backward looking, 
and predict clear Time-Reversal Invariance (TRI) violations for the volatility/square-return correlation function. 
Such violations are indeed observed empirically, although the magnitude of the effect is quite a lot smaller than predicted. 
This suggests that QARCH models, which postulate a {\it deterministic} relation between volatility and past returns, 
discard another important source of fluctuations. 
We postulate that ``the'' grand model should include both ARCH-like feedback effects and stochastic volatility effects, 
in such a way that TRI is only weakly broken. 
The stochastic volatility component could be the source of the extra kurtosis of the residuals 
noted above.\footnote{This discussion might be related to the interesting observation made by Virasoro in \cite{virasoro2011non}.}

The present paper is, to the best of our knowledge, the first attempt at unveiling the fine structure volatility feedback effects 
in autoregressive models. We believe that it is a step beyond the traditional econometric approach of postulating a convenient mathematical model,
which is then brute-force calibrated on empirical data. What we really need is to identify the {\it underlying mechanisms} that would justify, at 
a deeper level, the use of a QARCH family of model rather than any another one, for example the multifractal random walk model. 
From this point of view, we find remarkable that the influence of daily returns is so strongly singled out by our empirical results, when we 
expected that other time scales would emerge as well. The quandary lies in the unexpectedly complex structure of the off-diagonal 
feedback component, for which we have no interpretation. 

{A natural extension of our study, presented in the companion paper \cite{Pierre_inprep} that sheds further light on a
possible behavioral interpretation of volatility feedback is to decompose daily returns into higher frequency components 
(overnight and intraday returns, or 5 minute returns).}
In this view, the feedback mechanism of the volatility at the intraday scale might be related to the self-excitement of order arrivals in the book \cite{bacry2012non}.
Many other remaining questions should be addressed empirically, for example 
the dependence of the feedback effects on market capitalization, average volatility, etc. Another interesting generalisation would be 
a double regression of the volatility on the past returns of stocks and of those of the index. Finally, other financial assets, such 
as currencies or commodities, should be studied as well. Stocks, however, offer the advantage that the data is much more abundant, specially if 
one chooses to invoke some structural universality, and to treat all stocks as different realizations of the same underlying process.

%%% blinding
\paragraph{Acknowledgements}
We thank R.~Allez, P.~Blanc, M.~Potters and M.~Virasoro for useful discussions, 
and G.~Zumbach for his careful reading and numerous comments.

\appendix
\section{Fourth moment of the volatility in ARCH($q$)}\label{sec:fourth_moment}

In the general case, $\langle \sigma^4 \rangle$, $\mathcal{C}^{(2)}(\tau)$ and $\mathcal{D}(\tau',\tau'')$ 
are related by the following set of self-consistent equations:
\begin{subequations}
\begin{align}
	\vev{\sigma^4}-\vev{\sigma^2}^2 =&
	\vev{\sigma^2}^2\Big(\tr(K^2)-\tr(K^{\bullet 2})\Big)+
	   \sum_{\tau=1}^qK(\tau,\tau){\mathcal{C}}^{(2)}(\tau)\\\nonumber%
	&+2\sum_{\tau_1>\tau_2=1}^qK(\tau_1,\tau_2)\left[{\mathcal{D}}(\tau_1,\tau_2)-\sum_{\tau=1}^{\tau_2-1}K(\tau,\tau)\mathcal{D}(\tau_1-\tau,\tau_2-\tau)\right]\\
	{\mathcal{C}}^{(2)}(\tau>0)=&K(\tau,\tau)\Big(\vev{\sigma^4}\vev{\xi^4}-\vev{\sigma^2}^2\Big)+
	\sum_{\tau'\neq\tau}K(\tau',\tau')\mathcal{C}^{(2)}(\tau-\tau')\\\nonumber
	&+2\sum_{\tau'>\tau''=\tau+1}^qK(\tau',\tau'')\mathcal{D}(\tau'-\tau,\tau''-\tau)\\
	{\mathcal{D}}(\tau_1 >0,\tau_2>0)=&2K(\tau_1,\tau_2)\Big(\mathcal{C}^{(2)}(\tau_1-\tau_2)+\vev{\sigma^2}^2\Big)+\sum_{\tau'=1}^{\tau_2-1}K(\tau',\tau')\mathcal{D}(\tau_1-\tau',\tau_2-\tau')\\\nonumber
	&+2\sum_{\tau'>\tau_2,\tau'\neq\tau_1}K(\tau',\tau_2)\mathcal{D}(\tau'-\tau_2,|\tau_1-\tau_2|).
\end{align}
\end{subequations}
where we assume for simplicity here that the leverage effect is absent, i.e.\ $L(\tau) \equiv 0$, 
and $K^{\bullet 2}$ means the square of $K$ in the Hadamard sense (i.e.\ element by element). 
For a QARCH with maximum horizon $q$, we have thus a set of $1+q+q(q-1)/2$ linear equations for $\mathcal{C}^{(2)}(\tau\geq 0)$
that can be numerically solved for an arbitrary choice of the kernel $K$. 
These equations simplify somewhat in the case of a purely diagonal kernel $K(\tau,\tau')=k(\tau) \delta_{\tau,\tau'}$. 
One finds:
\begin{subequations}\label{eq:vol4}
\begin{align}
	\vev{\sigma^4}=&\vev{\sigma^2}^2+\sum_{\tau>0}k({\tau})\mathcal{C}^{(2)}(\tau)\\
	{\mathcal{C}}^{(2)}(\tau)=&
     k({\tau})\Big(\vev{\sigma^4}\vev{\xi^4}-\vev{\sigma^2}^2\Big)+
	\sum_{\tau'\neq\tau>0}k({\tau'})\mathcal{C}^{(2)}(\tau-\tau')
\end{align}
\end{subequations}
By substituting $\vev{\sigma^4}$, it is easy to explicit the linear system in matrix form $\nabla\,\mathcal{C}^{(2)}=S$ with
\begin{subequations}
\begin{align}\label{eq:nabla_elements}
	\nabla(\tau,\tau')&=\delta_{\tau{\tau'}}-\vev{\xi^4}k({\tau})k({\tau'})-\left[k({\tau-\tau'})+k({\tau+\tau'})\right]\\
	S(\tau)&= k({\tau})\vev{\sigma^2}^2\left(\vev{\xi^4}-1\right),
\end{align}
\end{subequations}
and the convention that $k({\tau})=0, \forall\tau\leq 0$.

Let us examine this in more detail for ARCH($q$). 
For simplicity, we assume here that $\xi$ is Gaussian ($\vev{\xi^4}=3$) and $s$ is chosen such that $\vev{\sigma^2}=1$. 
The condition on $k(\tau)$ for which $\vev{\sigma^4}$ diverges is given by $\det \nabla =0$, 
where $\nabla$ is the matrix whose entries are defined in Eq.~\ref{eq:nabla_elements}. 
For different $q$'s, this reads:
\begin{itemize}
\item{for $q=1$}, one recovers the well known result that ARCH(1) has a finite fourth moment only when \mbox{$k_1<1/\sqrt{3}$}.
\item{for $q=2$}, the stability line is given by $k_1 + k_2 = 1$, while the existence of a finite fourth moment is given by the condition \mbox{$k_1^2 < (1/3-k_2^2)(1-k_2)/(1+k_2)$}.
\item{for $q \to \infty$}, we again assume the $\tau$ dependence of $k(\tau)$ to be a power-law, $g\,\tau^{-\alpha}$ (corresponding to the FIGARCH model). 
                  The critical line for which the fourth moment diverges is shown in dashed blue in Fig.~\ref{fig:sig2crit}. 
                  After a careful extrapolation to $q = \infty$, we find that whenever \mbox{$1 < \alpha < \alpha_c \approx 1.376$}, 
		  the fourth moment exists as soon as the model is stationary, i.e.\ when \mbox{$g<1/\zeta(\alpha)<1/\zeta(\alpha_c)\approx 0.306$}. 
\end{itemize}
The last result is quite interesting and can be understood from Eq.~\eqref{eq:vol4}, which shows that to lowest order in $g$, one has:
\be\label{eq:sigma4pert}
\frac{\langle \sigma^4 \rangle}{\langle \sigma^2 \rangle^2} - 1 \approx (\vev{\xi^4}-1) g^2 \sum_{\tau > 0} \frac{1}{\tau^{2\alpha}}.
\ee
The above expression only diverges if \mbox{$\alpha < 1/2$}, 
but this is far in the forbidden region \mbox{$\alpha < 1$} where $\langle \sigma^2 \rangle$ itself diverges. 

\section{Power-law volatility correlations in FIGARCH}
 \label{QARCHapx:B}
We provide here a more precise insight on the behavior of 
the quadratic correlation $\mathcal{C}^{(2)}(\tau)$ when the input diagonal kernel $k(\tau)$ is long-ranged asymptotically power-law:
\[
    k(\tau)\xrightarrow{\tau\to\infty}g/\tau^{1+\epsilon}, \quad 0<\epsilon,
\]
where the bound on $\epsilon$ ensures the integrability of the kernel $\int_0^\infty k(\tau)\d{\tau}=1-s^2$.

In order to address this question analytically, we assume that the feedback kernel is infinitely-ranged, 
and consider the continuous-time approximation.
The sum in Eq.~\eqref{eq:vol4} is approximated by an integral, and decomposed as follows
\begin{align*}
    \mathcal{C}^{(2)}(\tau)=\mathcal{C}^{(2)}(-\tau)&=             \int_0^\infty    k(\tau')\mathcal{C}^{(2)}(\tau-\tau')\,\d{\tau'}\\
                    &= \underbrace{\int_0^\tau      k(\tau')\mathcal{C}^{(2)}(\tau-\tau')\,\d{\tau'}}_{\mathcal{C}^{(2)}_-(\tau)}+
                       \underbrace{\int_\tau^\infty k(\tau')\mathcal{C}^{(2)}(\tau'-\tau)\,\d{\tau'}}_{\mathcal{C}^{(2)}_+(\tau)}
\end{align*}

The behavior at large $\tau\to\infty$ is studied by taking the Laplace transform 
and investigating $\omega\to0$ while keeping a non-diverging product $\omega\tau$:
\begin{align*}
    \widehat{k}(\omega)&=\int_0^\infty 1-\left(1-\e^{-\omega\tau}\right)k(\tau)\,\d{\tau}
                    =\widehat{k}(0)-\int_0^\infty \left(1-\e^{-x}\right)k(x/\omega)\,\d{x}/\omega\\
                   &=\widehat{k}(0)-\int_0^\infty \left(1-\e^{-x}\right)g\left(\frac{\omega}{x}\right)^{1+\epsilon}\,\d{x}/\omega\\
                   &=\widehat{k}(0)-g\omega^{\epsilon}\int_0^\infty \left(1-\e^{-x}\right)x^{-1-\epsilon}\,\d{x}
                    =\widehat{k}(0)+g\omega^{\epsilon}\int_0^\infty \e^{-x}\,\frac{x^{-\epsilon}}{-\epsilon}\,\d{x} \\
                   &=\widehat{k}(0)+g\omega^{\epsilon}\frac{\Gamma(1-\epsilon)}{-\epsilon}=
                     \widehat{k}(0)+g\,\Gamma(-\epsilon)\omega^{\epsilon}
\end{align*}

%\clearpage
%\subsubsection*{Power-law resulting correlation}
Empirical observations motivate the following Ansatz for the quadratic correlation:
\[
    \mathcal{C}^{(2)}(\tau)\xrightarrow{\tau\to\infty}B/\tau^\beta,\quad 0<\beta<1.
\]
We hope to be able to reconciliate the ``fast'' decay of $k$ (since integrable)
with a very slow asymptotic decay of the solution, by finding a 
relationship between $\beta$ and $\epsilon$.
In Laplace space, we have
\begin{align*}
    \widehat{\mathcal{C}^{(2)}}(\omega)&=\int_0^\infty \e^{-\omega\tau}\,\mathcal{C}^{(2)}(\tau)\,\d{\tau}
                                        =\int_0^\infty \mathcal{C}^{(2)}(x/\omega)\,\e^{-x}\,\d{x}/\omega\\
                   &\xrightarrow{\omega\to0}\int_0^\infty B\left(\frac{\omega}{x}\right)^{\beta}\e^{-x}\frac{\d{x}}{\omega}
                                        =B\,\Gamma(1-\beta)\,\omega^{\beta-1}\\
    \widehat{\mathcal{C}^{(2)}_-}(\omega)&=\widehat{k}(\omega)\, \widehat{\mathcal{C}^{(2)}}(\omega)\\
                     &\xrightarrow{\omega\to0}B\,\Gamma(1-\beta)\,\omega^{\beta-1} \left[(1-s^2)-g\frac{\Gamma(1-\epsilon)}{\epsilon}\omega^{\epsilon}\right]\\
        {\mathcal{C}^{(2)}_+}(\tau)  &=\int_0^\infty k(\tau+u)\,\mathcal{C}^{(2)}(u)\,\d{u}\qquad\text{dominated by the large $u\sim\tau$}\\
                     &\xrightarrow{\tau\to\infty}\int_0^\infty \frac{B}{u^\beta}\frac{g}{(\tau+u)^{1+\epsilon}}\,\d{u}
                                                =\tau^{-\beta-(1+\epsilon)+1}\int_0^\infty \frac{B}{x^\beta}\frac{g}{(1+x)^{1+\epsilon}}\d{x}\\
                                               &=\tau^{-(\beta+\epsilon)}gB\int_0^1 \left(\frac{y}{1-y}\right)^\beta y^{1+\epsilon}\frac{\d{y}}{y^2}
                                                =\tau^{-(\beta+\epsilon)}gB\frac{\Gamma(\epsilon+\beta)\,\Gamma(1-\beta)}{\Gamma(1+\epsilon)}\\
    \widehat{\mathcal{C}^{(2)}_+}(\omega)&=\int_0^\infty\e^{-\omega\tau}\,\mathcal{C}^{(2)}_+(\tau)\,\d{\tau}=\int_0^\infty \mathcal{C}^{(2)}_+(x/\omega)\,\e^{-x}\,\d{x}/\omega\\
                     &\xrightarrow{\omega\to0}gB\frac{\Gamma(\epsilon+\beta)\,\Gamma(1-\beta)}{\epsilon\Gamma(\epsilon)}\,\omega^{\epsilon+\beta-1}\,\Gamma(1-\epsilon-\beta)
\end{align*}
Collecting all the terms, we finally get
\begin{align*}
    B\,\Gamma(1-\beta)\,\omega^{\beta-1}&=B\,\Gamma(1-\beta)\,\omega^{\beta-1} \left[(1-s^2)-g\frac{\Gamma(1-\epsilon)}{\epsilon}\omega^{\epsilon}\right]\\
                                    &+gB\,\frac{\Gamma(\epsilon+\beta)\,\Gamma(1-\beta)}{\epsilon\Gamma(\epsilon)}\,\omega^{\epsilon+\beta-1}\,\Gamma(1-\epsilon-\beta)\\
    s^2B\,\Gamma(1-\beta)\,\omega^{\beta-1}&=gB\,\omega^{\epsilon+\beta-1}\,\Gamma(1-\beta) \left[\frac{\Gamma(\epsilon+\beta)\,\Gamma(1-\epsilon-\beta)}{\epsilon\,\Gamma(\epsilon)}-\frac{\Gamma(1-\epsilon)}{\epsilon}\right]
\end{align*}
Very surprisingly, the functional dependence is perfectly equalized in the limit $s^2\to0$: 
the different terms have compatible powers of $\omega$.
For the relationship to be an equality, it is necessary that $s^2\to0$ 
(i.e.\ the model is at the critical limit of quadratic non-stationarity),
and simultaneously the RHS term vanishes:
\begin{align*}
    \Gamma(\epsilon+\beta)\,\Gamma(1-\epsilon-\beta)&=\Gamma(1-\epsilon)\,\Gamma(\epsilon)\\
    \sin\left(\pi(\epsilon+\beta)\right)&=\sin\left(\pi\epsilon\right)\\
    (\epsilon+\beta)-\epsilon=2n&\quad\text{or}\quad(\epsilon+\beta)+\epsilon=1+2n,\qquad n\in\mathds{Z}
\end{align*}
For $\widehat{k}(0)=1-s^2\to 1$ to hold, $\epsilon$ must be close to $0$, 
and in this case there is only one solution 
\[
    \boxed{\beta=1-2\epsilon, \qquad 0<\epsilon<\tfrac{1}{2}}.
\]

\section{Methodology}\label{apx:methodology}

\subsection{Dataset}

Equation \eqref{eq:quadraticARCH} is a prediction model for the predicted variable 
$\sigma_t$ with explanatory variables past returns $r$ at all lags. The dataset we will use 
to calibrate the model is composed of daily stock prices (Open, Close, High, Low) 
for $N=280$ names present in the S\&P-500 index from Jan.\ 2000 to Dec.\ 2009 ($T=2515$ days), without 
interruption. The reference price for day $t$ is defined to be the close of that day $C_t$, and the return 
$r_t$ is given by $r_{t} = \ln C_{t} - \ln C_{t-1}$. The true volatility is of course unobservable; we 
replace it by the standard Rogers-Satchell (RS) estimator \cite{rogers1991estimating,floros2009modelling}: 
\be
	\widehat \sigma_t^2=\ln(H_t/O_t)\ln(H_t/C_t)+\ln(L_t/O_t)\ln(L_t/C_t).
\ee
As always in this kind of studies over extended periods of time, 
our dataset suffers from a selection bias since we have retained only 
those stock names that have remained alive over the whole period.

There are several further methodological points that we need to specify here:
\begin{itemize}
\item {\it Universality hypothesis}. We assume that the feedback 
matrix $K$ and the leverage kernel $L$ are identical for all stocks, once returns are standardized to get rid of 
the idiosyncratic average level of the volatility. This will allow us to use the whole data set (of size $N \times T$) to calibrate the model. 
Some dependence of $K$ and $L$ on global properties of firms (such as market cap, liquidity, etc.) may be
possible, and we leave this for a later study. However, we will see later that the universality hypothesis appears to 
be a reasonable first approximation.

\item {\it Removal of the market-wide volatility}. We anticipate that the volatility of a single 
stock has a market component that depends on the return of the index, and an idiosyncratic 
component that we attempt to account for with the returns of the stock itself. As a proxy for
the instantaneous market volatility, we take the cross-sectional average of the squared returns of individual
stocks, i.e.
\be
\Sigma_t = \sqrt{\frac{1}{N} \sum_{j=1}^N r_{j,t}^2 }
\ee
and redefine returns and volatilities as $r_t/\Sigma_t$ and $\widehat \sigma_t/\Sigma_t$ --- in order to avoid artificially capping high returns,
we in fact compute $\Sigma_t$ for every stock $i$ separately by performing the sum on $j\neq i$. Finally, as 
announced above, the return time series are centered and standardized, and the RS volatility time series
are standardized such that $\langle \widehat \sigma_{i,t}^2 \rangle = 1$ for all $i$s. 
(This also gets rid of the multiplicative bias of the Rogers-Satchell estimator when used with non-Gaussian returns.)

\item {\it Choice of the horizon $q$}. In principle, the value of the farthest relevant lag $q$ is an additional free parameter 
of the model, and should be estimated jointly with all the others. However, this would lead to a huge computational effort 
and to questionable conclusions. In fact, we will find that the diagonal elements $K(\tau,\tau)$ decay quite slowly with 
$\tau$ (in line with many previous studies) and can be accurately determined up to large lags using the GMM.
Off-diagonal elements, on the other hand, turn out to be much smaller and rather noisy. We will therefore restrict 
the horizon for these off-diagonal elements to $q=10$ (two weeks) or $q=20$ (four weeks). Longer horizons, although possibly
still significant, lead to very small out-of-sample extra predictability (but note that longer horizons {\it are} needed for the
diagonal elements of $K$).
\end{itemize}

\subsection{GMM estimation based on correlation functions}\label{sec:first_est}

On top of the already defined four-point correlation functions $\mathcal{C}^{(2)}(\tau)$ and $\mathcal{D}(\tau',\tau'')$ (and their 
corresponding ``tilde'' twins), we will introduce two- and three-point correlation functions that turn out to be useful (note that
the $r_t$s are assumed to have zero mean):
\begin{subequations}
\begin{align}
	           \mathcal{C}^{(1)}(\tau)        &\equiv\vev{r_t r_{t-\tau}}_t\\
	           \mathcal{C}^{(a)}(\tau)        &\equiv\vev{\left(r^2_t-\vev{r^2}\right)|r_{t-\tau}|}_t\\
	\widetilde{\mathcal{C}}^{(a)}(\tau)       &\equiv\vev{\left(\sigma^2_t-\vev{\sigma^2}\right)|r_{t-\tau}|}_t\\
	           \mathcal{L}      (\tau)        &\equiv\vev{\left(r^2_{t}-\vev{r^2}\right)r_{t-\tau}}_t\\
	\widetilde{\mathcal{L}}     (\tau)        &\equiv\vev{\left(\sigma^2_{t}-\vev{\sigma^2}\right)r_{t-\tau}}_t\\
	           \mathcal{L}^{(a)}(\tau)        &\equiv\vev{      |r_t| r_{t-\tau}}_t\\
	           \mathcal{D}^{(a)}(\tau',\tau'')&\equiv\vev{\left(|r_t|-\vev{|r|}\right)r_{t-\tau'}r_{t-\tau''}}.
\end{align}
\end{subequations}
The $\mathcal{C}^{(1)}(\tau)$ correlation function is by definition equal to $\vev{r_t^2}_t=1$ for $\tau=0$, 
and is usually considered to be zero for $\tau > 0$. 
However, as is well known, there are small anti-correlations of stock returns. 
On our data set, we find that these linear correlations are very noisy but significant, and can be fitted by:
\be
\mathcal{C}^{(1)}(\tau \geq 1) \approx - 0.04 \, \e^{-0.39 \tau},
\ee
corresponding to a decay time of \mbox{$\approx 2.5$} days.
The values of $\mathcal{C}^{(a)}$ characterize volatility correlations and are similar in spirit to $\mathcal{C}^{(2)}$, but they only involve 
third order moments of $r$, instead of fourth order moments, and are thus more robust to extreme events. 
The $\mathcal{L}$ correlations, on the other hand, characterize the leverage effect, 
i.e.\ the influence of the {\it sign} of past returns on future volatilities. 

These correlation functions allow us to define a well-posed problem of solving a system with \mbox{$1+q+\frac{q(q+1)}{2}$} unknowns
$\left(s^2, L(\tau), K(\tau',\tau'')\right)$ using the following \mbox{$1+q+q+\frac{q(q-1)}{2}$} equations that involve empirically 
measured correlation functions (in calligraphic letters), for \mbox{$1\leq\tau\leq q$} and \mbox{$1\leq\tau_2<\tau_1\leq q$}:
\begin{subequations}\label{eq:model_predictions}
\begin{align}\label{eq:2points}
	\vev{\sigma^2}&=s^2+\sum_{\tau',\tau''}K(\tau',\tau'')\mathcal{C}^{(1)}(\tau'\!-\!\tau'')\\
	\widetilde{\mathcal{L}}(\tau)&=\sum_{\tau'}L(\tau')\mathcal{C}^{(1)}(\tau\!-\!\tau')
	                         + \sum_{\tau'}K(\tau',\tau')\mathcal{L}(\tau\!-\!\tau')
	                         +2\sum_{\tau'\neq \tau}K(\tau',\tau)\mathcal{L}(\tau'\!-\!\tau)\\\label{eq:3pointsC}
	\widetilde{\mathcal{C}}^{(a)}(\tau)&\approx\sum_{\tau'}L(\tau')\mathcal{L}^{(a)}(\tau'\!-\!\tau)+
	\sum_{\tau'}K(\tau',\tau')\mathcal{C}^{(a)}(\tau-\tau')\\\nonumber
	&+2\sum_{\tau'>\tau''>\tau>0}K(\tau',\tau'')\mathcal{D}^{(a)}(\tau'\!-\!\tau,\tau''\!-\!\tau)\\\label{eq:4points}
	\widetilde{\mathcal{D}}(\tau_1,\tau_2) &\approx L(\tau_2)\mathcal{L}(\tau_1\!-\!\tau_2)+L(\tau_1)\mathcal{L}(\tau_2\!-\!\tau_1)\\\nonumber
	                              &+2\sum_{\tau'>\tau_2}K(\tau',\tau_2)\left(\mathcal{D}(\tau_1\!-\!\tau_2,\tau'\!-\!\tau_2)
	                              +\mathcal{C}^{(1)}(\tau_1\!-\!\tau')-\mathcal{C}^{(1)}(\tau'\!-\!\tau_2)\mathcal{C}^{(1)}(\tau_1\!-\!\tau_2)\right)\\\nonumber
	                              &+\sum_{\tau'\leq\min(\tau_1,\tau_2)}K(\tau',\tau')\mathcal{D}(\tau_1\!-\!\tau',\tau_2\!-\!\tau'),
\end{align}
\end{subequations}
where all the sums only involve positive $\tau$s.
These equations are exact if all 3-point and 4-point correlations that involve $r$s at 3 (resp.\ 4) distinct times are strictly zero.
But since the linear correlations $\mathcal{C}^{(1)}(\tau>0)$ are very small, it is a safe approximation to neglect these higher order 
correlations. 

Note that the above equations still involve fourth order moments (the off-diagonal elements of $\mathcal{D}$),
that in turn lead to very noisy estimators of the off-diagonal of $K$.
In order to improve the accuracy of these estimators, 
we have cut-off large events by transforming the returns $r_t$ into $r_{\text{cut}} \tanh(r_t/r_{\text{cut}})$, 
which leaves small returns unchanged but caps large returns. 
We have chosen to truncate events beyond $3-\sigma$, i.e.\ $r_{\text{cut}}=3$. 
In any case, we will use the above equations in conjunction with maximum likelihood (for which no cut-off is used) to obtain more robust estimates of these off-diagonal elements.

\subsection{Error estimates of the MLE for $\mathcal{K}^*$}\label{ssec:errorMLE}
%RECUPERER LA METHODOLOGIE DE Sect.~\ref{ssec:GMMML}

%Before exposing our results, 
We briefly go through a discussion of the bias and error 
on the estimated parameters $\mathcal{K}^*$ as well as on the resulting maximal average likelihood $\overline{\mathcal{I}}^*$.
The likelihood $\mathcal{I}$, its gradient $\partial {\mathcal{I}}$ and Hessian matrix $\partial \partial {\mathcal{I}}$
are generic functions of the set of parameters $\mathcal{K}$ to be estimated, and of the dataset, of size $n$.
As the number $n$ of observations tends to infinity, the covariance matrix of the ML estimator of the parameters is well known to be $(nI)^{-1}$,
where $I$ is the Fisher Information matrix
\[
	I=\mathds{E}\left[-\partial \partial {\mathcal{I}}\right]\approx -\overline{\partial \partial {\mathcal{I}}}(\mathcal{K}^*)
	                                                        %\approx -\partial \partial {\mathcal{I}}(K^{\infty}),
\]
while the asymptotic bias scales as $n^{-1}$ and is thus much smaller than the above error ($\sim n^{-1/2}$). 
As a consequence, ML estimates of $\mathcal{K}$ exceeding $\pm \operatorname{diag}(-n\,\overline{\partial\partial\mathcal{I}}^*)^{-1/2}$ will be deemed significant.
The {\it bias} on the average in-sample (IS) value of the maximum likelihood itself can be computed to first order in $1/n$, 
and is very generally found to be $+M/2n$, where $M$ is the number of parameters to be determined.
Similarly, the bias on the average out-of-sample value of the maximum likelihood is $-M/2n$.%
These corrections to the likelihoods 
lead to the (per point) Akaike Information Criterion \cite{akaike1974new} \mbox{$AIC=-2(\mathcal{I}-M/n)$}, 
that trades off the log-likelihood and the number of parameters. AIC is used for model selection purposes mainly. 
When comparing parametric models with the same number of parameters, AIC is not more powerful than the likelihood.

Since each sampling (half the pool of stocks) of our data set  contains $n = T\cdot N/2\approx 350,000$ observations, 
differences of likelihood smaller than $M/2n \approx 3 \cdot 10^{-4}$ are 
insignificant when $M=190$ (corresponding to all off-diagonal elements when $q=20$). 
This number is $\approx 5$ times smaller when one considers the restricted models introduced in Section 2 (which contain $\approx 40$ parameters).

\section{Empirical study: stock index}\label{sec:emp_index}
 \begin{figure}
 	\center
 	\includegraphics*[scale=0.6,angle=-90]{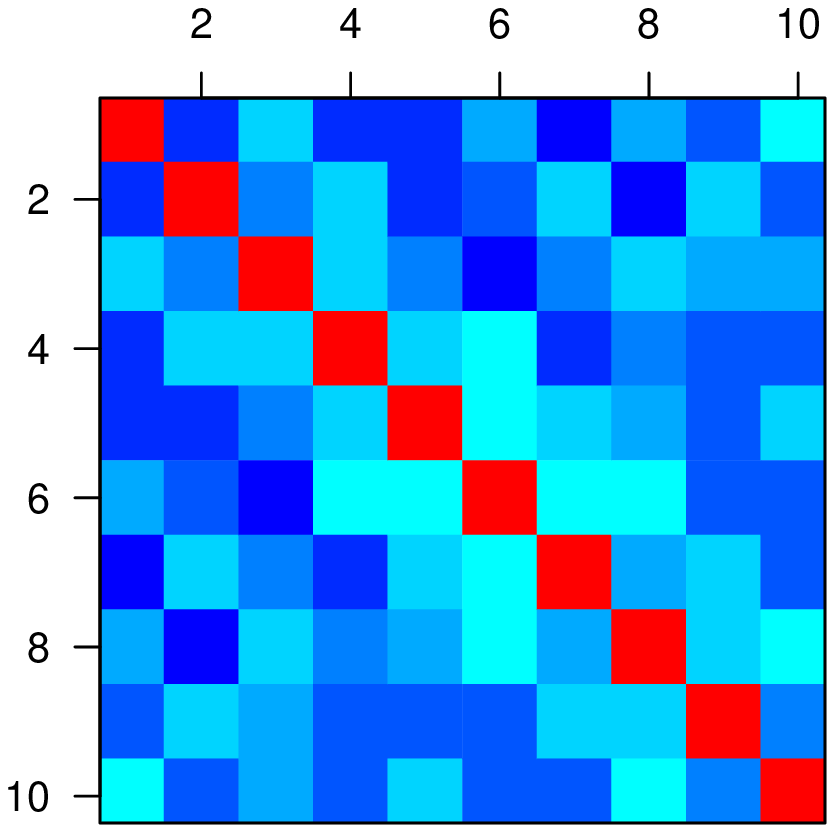}
 	\includegraphics*[scale=0.6,angle=-90]{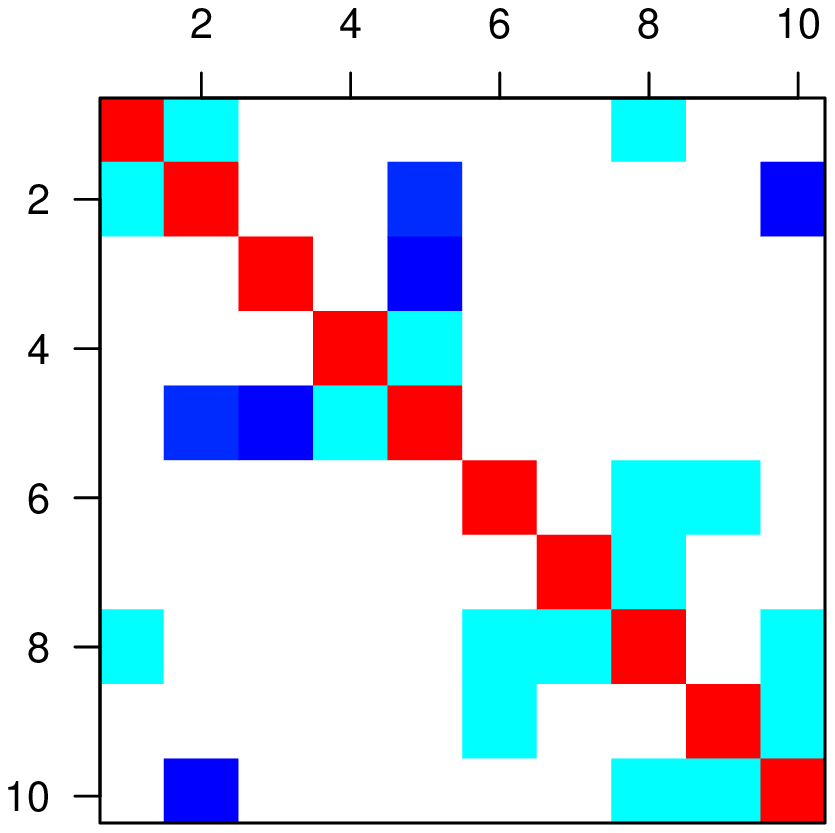}
 	\includegraphics*[scale=0.35,angle=-90]{colkeys.eps}
 	\caption{Heatmap of the $q=10$ kernel for the index volatility.
 	\textbf{Left: }the GMM estimation; \textbf{Right: }the ML estimation with GMM prior 
 	(again, we have checked that the ARCH prior leads to very close results).}
 	\label{fig:heatmap_idx}
 \end{figure}

We complete our analysis by a study of the returns of the  S\&P-500 index in the QARCH framework.
We use a long series of more than 60 years, from Oct.\ 1948 to Sept.\ 2011 ($15\,837$ days).

The computation of the correlation functions and the GMM calibration of a long ARCH(512) yields a $s^2(q)$ 
that can be fitted with the formula~\eqref{eq:fit_s2} and the following parameters: 
$s_{\infty}^2\approx0.20, \alpha\approx 1.28, g\approx 0.162$ and $q_0\approx262$, yielding fluctuations $\vev{\sigma^4}\approx 1.688$.
Contrarily to single stocks, the One-step Maximum Likelihood estimation of the QARCH failed with $q=20$,
as the gradient and Hessian matrix evaluated at the arrival point are, respectively, large and not negative definite. 
Although the starting point appears to be close to a local maximum (the Hessian matrix is negative definite), 
the one-step procedure does not lead to that maximum.

In order to control better the Maximum Likelihood estimation, we lower the dimensionality of the parameter space 
and estimate a QARCH(10), although still with a long memory diagonal ($q=50$).
Here the procedure turns out to be legitimate, and the resulting kernel $K$ is depicted in Fig.~\ref{fig:heatmap_idx} (right). 
Interestingly, the off-diagonal content in the QARCH model for index returns is mostly not significant 
(again, white regions correspond to values not exceeding their theoretical uncertainty) 
apart from a handful of negative values around $\tau=8-10$ and one row/column at $\tau=5$. 
The contribution of the latter to the QARCH feedback can be made explicit as
\[
	2\,r_{t-5}\sum_{\tau<5}K(5,\tau)r_{t-\tau}
\]
and describes a trend effect between the daily return last week $r_{t-5}$ and the (weighted) cumulated return since then 
$\sum_{\tau<5}K(5,\tau)r_{t-\tau}$. 
It would be interesting to know whether this finding is supported by some intuitive feature of the trading activity on the S\&P-500 index. 
Note that, again, none of the ``simple'' structures discussed in Sect.~\ref{sec:section3} is able to account for the structure of $K^*$
(compare Fig.~\ref{fig:heatmap_idx} with Fig.~\ref{fig:matrices}).

The spectral analysis reveals a large eigenvalue much above the ARCH prediction, and a couple of very small eigenvalues,
similarly to what was found for the stock data.
However, the eigenvectors associated with them exhibit different patterns: 
the first eigenvector does not reveal the expected collective behavior, but rather a dominant $\tau=1$ component, 
with a significant $\tau=2$ component of opposite sign.
The other modes do not show a clear signature and are hard to interpret.

The procedure for computing in-sample/out-of-sample likelihoods is similar to the case of the stock data, 
but the definitions of the universes differs somewhat since we only have a single time series at our disposal. 
Instead of randomly selecting half of the series, we select half of the dates (in block, to avoid breaking the time dependences) 
to define the in-sample universe $\Omega$,
on which the correlation functions are computed and both estimation methods (GMM, and one-step maximum likelihood) are applied.
Then we evaluate the likelihoods of the calibrated kernels, first on $\Omega$ to obtain the `in-sample' likelihoods, 
and then on the complement of $\Omega$ to get the `out-of-sample' likelihoods.
We iterate $N_{\text{samp}}=150$ times and draw a random subset of dates every time, then average the likelihoods, 
that we report in the Table~\ref{tab:IS-OOS.idx}. 
The 1-s.d.\ systematic dispersion of the samplings is now $\approx 7 \cdot 10^{-3}$.
Because of the fewer number of observations in the index data compared to the stock data, 
corrections for the bias induced by the number of parameters $M$ become relevant. 
Adjusting the out-of-sample likelihood by subtracting the bias $-M/2n \approx 3 \cdot 10^{-3}$ (with $n\approx T/2=7.5\cdot 10^3$ and $M=q(q-1)/2=45$), 
brings the ARCH+ML result to a level competitive with ARCH (but not obviously better), and certainly better than the GMM estimate.

\begin{table}
\begin{center}
\begin{tabular}{c||cc|c|}
  &GMM        &ARCH(10)   &ARCH+ML     \\\hline\hline	%&ARCH+ML+veto
IS&$-1.16750$ &$-1.16666$ &${-1.16522}$\\		%&-1.16487
OS&$-1.16972$ &$-1.16704$ &${-1.17079}$\\\hline		%&-1.17066
\end{tabular}
\caption{Average log-likelihoods, according to Eq.~\eqref{eq:loglike}, for the stock index.
}
\label{tab:IS-OOS.idx}
\end{center}
\end{table}

\nocite{cont2010encyclopedia}
\nocite{berd2011lessons}
\nocite{engle1986handbook}
\nocite{teyssiere2010long}
\bibliographystyle{abbrv}
\bibliography{../biblio_all}

\end{document}